\documentclass[a4paper,11pt]{article}
\pdfoutput=1

\usepackage{jheppub}
\usepackage[T1]{fontenc}
\usepackage[utf8]{inputenc}
\usepackage{microtype}
\usepackage{lmodern}
\usepackage{booktabs}
\usepackage{mdframed}
\usepackage{graphicx}
\usepackage{verbatim}
\usepackage{hyperref}
\usepackage{ulem}
\usepackage{xcolor}
\definecolor{dark-red}{rgb}{0.4,0.15,0.15}
\definecolor{dark-blue}{rgb}{0.15,0.15,0.4}
\definecolor{medium-blue}{rgb}{0,0,0.5}
\hypersetup{colorlinks, linkcolor={dark-red}, citecolor={dark-blue}, urlcolor={medium-blue}}

\usepackage{tikz}
\usetikzlibrary{arrows.meta, decorations.pathmorphing, positioning}

\usepackage{amsmath,amsfonts,amssymb}
\usepackage{eucal}
\usepackage{mleftright}
\usepackage{tensor}
\usepackage{braket}
\usepackage{bm}
\usepackage{mathtools}
\usepackage{mathrsfs}

\newcommand{\mL}{\mathcal{L}}
\newcommand{\mH}{\mathcal{H}}
\newcommand{\mK}{\mathcal{K}}

\newcommand{\mE}{\mathcal{E}}
\newcommand{\mO}{\mathcal{O}}

\newcommand{\dd}{\mathop{}\!\mathrm{d}}

\newcommand{\pilo}{\Pi} 
\newcommand{\glo}{h} 

\title{Kerroll black holes}

\author[a,b]{Florian Ecker,}
\author[a,b]{Daniel Grumiller,}
\author[a,b,c]{Lucas H{\"o}rl,}
\author[a,d]{Matt{\'e}o Leturcq--Daligaux,}
\author[b,e,f]{and Alfredo P{\'e}rez}

\affiliation[a]{Institute for Theoretical Physics, TU Wien, Wiedner Hauptstrasse 8-10, 1040 Vienna, Austria}
\affiliation[b]{Erwin Schr\"odinger International Institute for Mathematics and Physics, Boltzmanngasse 9, A-1090 Vienna, Austria}
\affiliation[c]{Department of Physics, ETH Zürich, 8093 Zürich, Switzerland}
\affiliation[d]{{\'E}cole normale sup{\'e}rieure --- PSL, 24 rue Lhomond, 75005 Paris, France}
\affiliation[e]{Centro de Estudios Cient{\'i}ficos (CECs), Avenida Arturo Prat 514, Valdivia, Chile}
\affiliation[f]{Facultad de Ingenier{\'i}a, Universidad San Sebasti{\'a}n, sede Valdivia, General Lagos 1163, Valdivia 5110693, Chile}

\emailAdd{fecker@hep.itp.tuwien.ac.at, grumil@hep.itp.tuwien.ac.at, lhoerl@ethz.ch, matteo.leturcq--daligaux@ens.psl.eu, alfredo.perez@uss.cl}

\abstract{%
We construct rotating black holes in Carroll gravity using two distinct approaches. In one of them, we exploit the freedom in the Carroll compatible connection to encode rotation. In particular, we construct rotating solutions in magnetic Carroll gravity by dressing the Carroll--Schwarzschild black hole with a rotational charge. This solution is intrinsically Carrollian and has no Lorentzian analog. In the other approach, we construct an extension of magnetic Carroll gravity from general relativity in an odd-power expansion in the speed of light. This theory contains magnetic Carroll gravity as a subsector but has in general more physical degrees of freedom. We show that this theory admits a Carroll analog of the Kerr black hole as a solution, which we refer to as the ``Kerroll black hole''. Its rotation appears as an intrinsically odd-power effect in the Carroll data. We compute the corresponding conserved charges for both theories.}
\begin{document} 
\maketitle


\section{Introduction\label{sec:intro}}
The Carroll limit formally arises as a vanishing speed of light contraction of the Poincaré algebra \cite{Levy1965,SenGupta1966OnAA} and has recently received a lot of attention, particularly in the context of flat space holography \cite{Barnich:2006av, Bagchi:2016bcd, Ciambelli:2018wre, Gupta:2020dtl, Donnay:2022aba, Bagchi:2022emh, Figueroa-OFarrill:2022mcy, Bagchi:2023cen, Poulias:2025eck, Fiorucci:2025twa}, see also \cite{Bagchi:2025vri, Ruzziconi:2026bix} for reviews. Additionally, Carroll symmetries have been found to emerge in areas such as cosmology \cite{deBoer:2021jej,deBoer:2023fnj}, tensionless strings \cite{Bagchi:2015nca,Bagchi:2016yyf,Bagchi:2020ats,Bagchi:2022iqb,Bagchi:2024qsb}, fractons \cite{Bidussi:2021nmp,Marsot:2022imf,Baig:2023yaz,Figueroa-OFarrill:2023vbj,Figueroa-OFarrill:2023qty,Pena-Benitez:2023aat,Ahmadi-Jahmani:2025iqc} and black holes \cite{Penna:2018gfx,Donnay:2019jiz,Redondo-Yuste:2022czg,Gray:2022svz,Bicak:2023rsz,Ecker:2023uwm,Bagchi:2023cfp,Bagchi:2024rje}, among many other applications \cite{Bagchi:2025vri}. In their most basic form, Carroll spacetimes are smooth manifolds equipped with a degenerate metric $h_{\mu \nu }$ of signature $(0,+,\cdots,+)$, together with a vector field $v^\mu $ that lies in its kernel, $h_{\mu\nu}\,v^\nu=0$. Upon gauging Carroll symmetries, various models of Carroll gravity have been developed, either as toy models or as limits of general relativity \cite{Henneaux:1979vn,Hartong:2015xda,Matulich:2019cdo,Grumiller:2020elf, Perez:2021abf, Hansen:2021fxi,deBoer:2021jej,Perez:2022jpr,Baiguera:2022lsw,Bergshoeff:2024ilz,Campoleoni:2022ebj,Musaeus:2023oyp,Grumiller:2024dql,Pekar:2024ukc,Aviles:2025ygw}. They give rise to dynamics for the metric variables $(h_{\mu \nu },v^\nu )$.

In this context, it has been shown that magnetic Carroll gravity, whose construction we shall review in Section \ref{sec:pre}, admits massive solutions \cite{Grumiller:2017sjh,Guerrieri:2021cdz, Perez:2021abf,Hansen:2021fxi,deBoer:2023fnj,Tadros:2024bev}. They were identified as Carroll black holes characterized by a Carroll extremal surface and Carroll thermal properties \cite{Ecker:2023uwm}. See also \cite{Aviles:2025ygw} for a more recent investigation of the latter in the context of $3$d Carroll gravity. A simple example is the Carroll--Schwarzschild black hole, defined by the Carroll metric data
\begin{align}
    v = v^\mu \partial_\mu =- \frac{1}{\sqrt{1-r_\mathrm{s}/r}}\,\partial_t
    \qquad\qquad
    h_{\mu\nu}\, \dd x^\mu \dd x^\nu
    = \frac{\dd r^2}{1-r_\mathrm{s}/r} + r^2\, \dd\Omega^2\,.
\end{align}
The Carroll extremal surface, i.e. the locus where $v$ becomes singular, coincides with the bifurcation sphere of the parent Lorentzian Schwarzschild black hole. The radius $r_\mathrm{s}=2G_\mathrm{M} E$ is given in terms of the rescaled (``magnetic'') Newton constant $G_\mathrm{M}=G_\mathrm{N}c^{-4}$ and the energy by $E=Mc^2$ with the mass $M$. Both $E$ and $G_\mathrm{M}$ remain fixed in the $c\to 0$ limit \cite{Ecker:2024czh}. 

So far, Carroll black hole constructions have essentially been restricted to static non-rotating solutions. As far as rotating geometries are concerned, there have been some recent investigations to recover rotation in a small-$c^2$-expansion \cite{Bal:2026xup}, but the strict Carroll limit leads to difficulties. One way to see this is by taking a limit of the Kerr metric in Boyer--Lindquist coordinates. Writing the Lorentzian metric in terms of $E$ and $G_\mathrm{M}$ yields
\begin{multline}
    \dd s^2 = -\Big(1-\frac{2EG_\mathrm{M}r}{\Sigma }\Big)c^2\dd t^2 -\frac{4EG_\mathrm{M}ar\sin ^2\theta }{\Sigma }c\,\dd t \dd \varphi +\frac{\Sigma }{\Delta }\dd r^2+\Sigma \dd \theta ^2\\
    +\Big(r^2+a^2+\frac{2EG_\mathrm{M}a^2r\sin ^2\theta }{\Sigma }\Big)\sin ^2\theta \dd \varphi ^2 \label{eq:kerr_mag_metric}
\end{multline}
where $\Delta = r^2-2G_\mathrm{M}Er+a^2$ and $\Sigma =r^2+a^2\cos ^2\theta $. The angular momentum parameter $a$ is related to the angular momentum charge $J$ by $a=Jc/E$, and one observes two simple ways of taking the limit: Either we fix $J$, implying $a\to 0$, thus reducing the geometry to the Carroll--Schwarzschild spacetime, or we fix $a$, which yields a diagonal Carroll metric 
\begin{align}
    h_{\mu \nu }\dd x^\mu \dd x^\nu =\frac{\Sigma }{\Delta }\dd r^2+\Sigma \dd \theta ^2+\Big(r^2+a^2+\frac{2EG_\mathrm{M}a^2r\sin ^2\theta }{\Sigma }\Big)\sin ^2\theta \dd \varphi ^2 ~.
\end{align}
However, such a metric cannot be a solution of magnetic Carroll gravity unless $a=0$. Indeed, in an ADM split the spatial metric is given by $g_{ij}=\delta _i^\mu \delta _j^\nu h_{\mu \nu }$ and should satisfy the constraint $R[g_{ij}]\approx 0$ with $R$ being the spatial Ricci scalar (see also Subsection \ref{subsec:mag}). One may readily check that this is not fulfilled for $a\neq 0$. Thus, there is no obvious Carroll limit of the Kerr spacetime that is a solution of magnetic Carroll gravity and at the same time carries non-zero angular momentum.
This is supported by a recent theorem \cite{Kolar:2025ebv} stating that every stationary axisymmetric solution of magnetic Carroll gravity in $D>3$ spacetime dimensions has to be static. While we agree with this theorem, we would like to argue that it does not exclude rotating geometries. Indeed, the notion of staticity\footnote{See, e.g., Wald \cite{waldgeneral}: Staticity is defined by the existence of a hypersurface orthogonal timelike Killing vector field. Alternatively, one could define staticity as stationarity plus time-reversal invariance.} used by \cite{Kolar:2025ebv} refers to the metric variables but is agnostic about the connection, which can carry independent geometric data that may spoil the invariance of spacetime under time-reversal. Therefore, in the Carrollian setting stationarity plus time-reversal symmetry are not equivalent to staticity since a spacetime can be static in the above sense but not time-reversal symmetric.

\enlargethispage{1truecm}

In this article, we show two ways of describing rotating Carroll black holes, circumventing the no-go theorem in \cite{Kolar:2025ebv}:
\begin{enumerate}
\item From a canonical point of view, asymptotic symmetry analyses in magnetic Carroll gravity \cite{Perez:2021abf,Perez:2022jpr} show that, for suitable boundary conditions, the surface charges form the Carroll algebra and in particular admit a non-vanishing $\mathfrak{so}(3)$ sector that may naturally be interpreted as describing angular momentum. Thus, non-zero angular momentum is in principle compatible with magnetic Carroll gravity. As we shall make more precise, it is the Carroll compatible connection that carries the relevant degrees of freedom for switching on these angular momentum charges. In a Hamiltonian language these degrees of freedom are given by the canonical momenta $\pi ^{ij}$ which, unlike for general relativity, cannot be solved for in terms of the other ADM variables and thus have to be treated as independent data. We shall show that within standard magnetic Carroll gravity, one can indeed construct rotating solutions by exploiting this freedom. Notably, these solutions are an intrinsically Carrollian construction without a known analog in Einstein gravity, in contrast to the Carroll black hole solutions constructed so far \cite{Ecker:2023uwm,Tadros:2024bev}.
\item The other approach introduces a new Carrollian gravitational theory obtained from an odd-power expansion in $c$. Here we start from the ADM action of general relativity and allow odd powers of $c$ in the expansion of all fields, thereby generalizing the so far used expansions in terms of only even powers \cite{Hansen:2021fxi,Ecker:2023uwm}. 
We then truncate the $\mathcal{O}(c^2)$-theory in a self-consistent way to find an extension of magnetic Carroll gravity. This theory, which we refer to as ``extended magnetic Carroll gravity'', may accommodate the rotational information of the Kerr metric in the fields that arise at odd powers of $c$ in the expansion. This provides a natural explanation for why purely even-power expansions fail to produce rotating Carroll black holes as a limit of the Lorentzian parent theory.
\end{enumerate}

The paper is organized as follows. In Section~\ref{sec:pre}, we review Carroll gravity in the Hamiltonian formulation by means of the two prominent examples of electric and magnetic Carroll gravity. In Section~\ref{sec:3}, we restrict to magnetic Carroll gravity and first summarize the chosen boundary conditions and asymptotic symmetries in Subsections \ref{sec:BCs} and \ref{sec:ASS}. We proceed by constructing rotating versions of flat Carroll spacetime and  Carroll--Schwarzschild by switching on an appropriate angular momentum charge in Subsections \ref{sec:flat_rot} and \ref{sec:CSS}. Finally, we analyze the respective bulk geometries, compute Killing fields and geodesics in Subsection \ref{sec:bulk_interp}. In Section~\ref{sec:4}, we develop the odd-power expansion of general relativity in ADM variables and define extended magnetic Carroll gravity in Subsection \ref{subsec:4.1}. We then present a two-parameter family of solutions to this theory, so-called Kerroll black holes, in Subsection \ref{sec:Kerroll_BHs}, and finally show in Subsection \ref{subsec:bdy-charges} that they carry a non-vanishing boundary charge associated with angular momentum. We conclude in Section~\ref{Conclusion} with a summary and outlook. 


\section{Carroll gravity}\label{sec:pre}
Carrollian limits of Einstein gravity are conveniently discussed in the Hamiltonian formulation \cite{Henneaux:2021yzg,Campoleoni:2022ebj}. Denoting the spatial metric as $g_{ij}$ and the conjugate momentum as $\pi^{ij}$, the action of general relativity in $d+1$ dimensions may be written as
\begin{align}\label{eq:GR_ADM}
    I[g_{ij},\pi^{ij},N,N^i]
    = \int \dd t \,\dd^d x \,
    \bigl(\pi^{ij}\dot{g}_{ij}-N\mathcal{H}-N^i\mathcal{H}_i\bigr) 
\end{align}
where $N$ is the lapse function and $N^i$ the shift vector. Hamiltonian and momentum constraints read
\begin{align}
    \mathcal{H}
    &= \frac{16\pi G_\mathrm{N}}{\sqrt{g}c^2}
        \Bigl(\pi^{ij}\pi_{ij}-\frac{1}{d-1}\pi^2\Bigr)
       -\frac{\sqrt{g}c^4}{16\pi G_\mathrm{N}}\,R[g] &
    \mathcal{H}_i
    &= -2\nabla_j\pi^j{}_i 
\end{align}
where $R[g]$ denotes the Ricci scalar of $g_{ij}$ and $\nabla_i$ its associated Levi-Civita connection. 

By virtue of the fundamental equal-time Poisson bracket $\{g_{ij}(x),\pi ^{kl}(y)\}=\delta _{(i}^k\delta _{j)}^l\delta ^d(x-y)$, the canonical Poisson brackets of the first-class constraints 
\begin{subequations}\label{eq:Poin_constraints}
\begin{align}
    \{\mathcal{H}(x),\,\mathcal{H}(y)\}&=c^2(\mH^i(x)+\mH^i(y))\,\partial _i \delta(x-y)\\
    \{\mathcal{H}(x),\,\mathcal{H}_i(y)\}&=\mathcal{H}(y)\,\partial _i \delta(x-y) \\
    \{\mathcal{H}_i(x),\,\mathcal{H}_j(y)\}&=\mathcal{H}_i(y)\,\partial _j \delta(x-y)
    +\mathcal{H}_j(x)\,\partial _i \delta(x-y)
\end{align}
\end{subequations}
upon taking the limit $c\to 0$ yield the zero-signature limit of the surface-deformation algebra \cite{Henneaux:1979vn},
\begin{subequations}\label{eq:Carr_constraints}
\begin{align}
    \{\mathcal{H}(x),\,\mathcal{H}(y)\}&=0\\
    \{\mathcal{H}(x),\,\mathcal{H}_i(y)\}&=\mathcal{H}(y)\,\partial _i \delta(x-y) \\
    \{\mathcal{H}_i(x),\,\mathcal{H}_j(y)\}&=\mathcal{H}_i(y)\,\partial _j \delta(x-y)
    +\mathcal{H}_j(x)\,\partial _i \delta(x-y)\,.
\end{align}
\end{subequations}
Importantly, this makes the right hand side of the first bracket vanish which is a key signature of Carrollian theories and responsible for an infinite tower of (supertranslation) charges. 

On the level of the action, there are two main possibilities to take the limit which differ in their choice of scaling of $G_\mathrm{N}$ as $c\to 0$. We shall briefly outline these two in the following. In each case, the action takes again the form \eqref{eq:GR_ADM}.

\subsection{Magnetic Carroll gravity}\label{subsec:mag}
Applying the rescaling $G_\mathrm{N}=c^4G_\mathrm{M}$, the $c\to 0$ limit of general relativity leads to magnetic Carroll gravity \cite{Bergshoeff:2017btm,Henneaux:2021yzg,Hansen:2021fxi,Campoleoni:2022ebj}. In this case, the constraints reduce to
\begin{align}\label{eq:constraints}
    \mathcal{H}=-\frac{\sqrt{g}}{16\pi G_\mathrm{M}}\,R[g]
    &&
    \mathcal{H}_i=-2\nabla _j\pi ^j{}_i 
\end{align}
which satisfies the Carrollian surface deformation algebra \eqref{eq:Carr_constraints}. The canonical momenta $\pi^{ij}$ in this case act as Lagrange multipliers for the constraint
\begin{align}\label{eq:Keq0}
    \dot{g}_{ij}-2\nabla _{(i}N_{j)}=2NK_{ij}\approx 0
\end{align}
setting the extrinsic curvature of the spatial hypersurfaces\footnote{We define $K_{ij}=-\frac12 \mathcal{L}_vh_{ij}$ where $v^\mu =(-1/N,N^i/N)$ is the Carroll vector field and $h_{ij}$ are the spatial components of a $d+1$-dimensional Carroll metric.} to zero. The other dynamical equation is obtained by varying with respect to $g_{ij}$ and reads
\begin{align}
    \dot{\pi }^{ij}&=-\frac{N\sqrt{g}}{16\pi G_\mathrm{M}}\Big(R^{ij}-\frac12 g^{ij}R\Big)+\frac{\sqrt{g}}{16\pi G_\mathrm{M}}\Big(\nabla ^i\nabla ^j N-g^{ij}\nabla ^2N\Big)\nonumber\\
    &\quad \quad +\nabla _k\big(N^k\pi ^{ij}\big)-\pi ^{kj}\nabla _kN^i-\pi ^{ki}\nabla _kN^j ~. \label{eq:pidot}
\end{align}
Therefore, for a geometry to be a solution to magnetic Carroll gravity, we need the four sets of equations \eqref{eq:constraints}, \eqref{eq:Keq0} and \eqref{eq:pidot} to be satisfied. Even though magnetic Carroll gravity enforces $K_{ij}=0$, it allows for non-trivial spatial curvature and massive solutions known as Carroll black holes \cite{Ecker:2023uwm}.

\subsection{Electric Carroll gravity}
Prescribing instead the scaling $G_\mathrm{N}=c^2G_\mathrm{E}$, one obtains electric Carroll gravity \cite{Isham:1975ur,Teitelboim:1978wv,Henneaux:1979vn,Hansen:2021fxi,Pekar:2024ukc} which is described by the set of constraints 
\begin{align}\label{eq:electricH}
    \mathcal{H}
    &=\frac{16\pi G_\mathrm{E}}{\sqrt{g}}\Big(\pi^{ij}\pi_{ij}
      -\frac{1}{d-1}\pi^2\Big) &&
    \mathcal{H}_i
    =-2\nabla_j \pi^j{}_i ~.
\end{align}
The constraints again obey the Carroll surface deformation algebra \eqref{eq:Carr_constraints} and the dynamical equations take the form
\begin{align}
    \dot g_{ij}
    &=\frac{32\pi G_\mathrm{E} N}{\sqrt{g}}\big(\pi_{ij}
      -\frac{1}{d-1}g_{ij}\pi\big)
      +\nabla_i N_j + \nabla_j N_i \\
    \dot\pi^{ij}
    &=\frac{8\pi G_\mathrm{E} N}{\sqrt{g}}\,g^{ij}
      \big(\pi^{kl}\pi_{kl}-\frac{1}{d-1}\pi^2\big)
      -\frac{32\pi G_\mathrm{E} N}{\sqrt{g}}\big(\pi^{i}{}_{k}\pi^{kj}
      -\frac{1}{d-1}\pi\,\pi^{ij}\big) \nonumber\\
    &\quad
      +\nabla_k\!\big(N^k\pi^{ij}\big)
      -\pi ^{kj}\nabla _kN^i-\pi ^{ki}\nabla _kN^j\,.
\end{align}
In contrast to the magnetic case, the Hamiltonian constraint is now purely kinetic and does not involve $R[g]$. The relation between $\pi^{ij}$ and the extrinsic curvature $K_{ij}$ is inherited from general relativity and is no longer just a constraint. Electric Carroll gravity can be interpreted as the strong-coupling or zero-signature limit of general relativity originally discussed in \cite{Isham:1975ur,Teitelboim:1978wv,Henneaux:1979vn}. It does not admit massive solutions \cite{Perez:2021abf,Hansen:2021fxi}.


\section{Rotating geometries in magnetic Carroll gravity}\label{sec:3}
Magnetic Carroll gravity admits massive solutions \cite{Perez:2021abf,Hansen:2021fxi} with an interpretation as Carroll black holes \cite{Ecker:2023uwm}. However, so far, this discussion has been restricted either to two spacetime dimensions or to the spherically symmetric sector in higher dimensions (with the exception of Kaluza--Klein reduced Carroll--BTZ black holes \cite{Ecker:2023uwm}). The possibility of having rotating Carroll black holes, i.e., solutions to magnetic Carroll gravity that are not static was recently excluded in \cite{Kolar:2025ebv}. 

In this Section, we reconsider this issue and point out a loophole in the above no-go result. In general, for theories of gravity rotation may be described by the presence of boundary charges that form a representation of some kinematical Lie algebra (see, e.g., \cite{Bergshoeff:2022eog}), such as the Carroll algebra. If the charges associated to the $\mathfrak{so}(d)$-part of this algebra are non-vanishing for a specific solution one may regard this as a first hint of non-trivial rotational properties. To make this more explicit we shall look at magnetic Carroll gravity in the following and show that such charges can indeed be present, even for asymptotically flat vacuum spacetimes that do not describe a black hole. 

To set the stage, we start with a brief summary of the chosen boundary conditions and asymptotic symmetries following \cite{Perez:2021abf}. From now on the analysis is restricted to four spacetime dimensions and we set $16\pi G_\mathrm{M}=1$ for the remainder of this Section.

\subsection{Boundary conditions and residual transformations}\label{sec:BCs}
Consider a foliation of spacetime by constant-time hypersurfaces that carry a spherical coordinate system $x^i=\{r,x^A\}$ with $x^A=\theta ,\varphi$. We impose boundary conditions for the ADM metric $g_{ij}$ and its momentum $\pi ^{ij}$ at $r\to\infty$, 
\begin{subequations}
\begin{align}
    g_{rr}&=1+\frac{f_{rr}}{r}+\frac{f_{rr} ^{(-2)}}{r^2}+\mathcal{O}(r^{-3}) \label{eq:bc1}\\
    g_{rA}&=\frac{f_{rA}^{(-1)}}{r}+\mathcal{O}(r^{-2})\\
    g_{AB}&=r^2\gamma _{AB}+rf_{AB}+f_{AB}^{(0)}+\mathcal{O}(r^{-1}) \phantom{\frac{1}{2}}\\
    \pi^{rr}&=p^{rr}+\mathcal{O}(r^{-1}) \phantom{\frac{1}{2}}\\
    \pi^{rA}&=\frac{p^{rA}}{r}+\frac{p^{rA}_{(-2)}}{r^2}+\mathcal{O}(r^{-3})\\
    \pi^{AB}&=\frac{p^{AB}}{r^2}+\mathcal{O}(r^{-3}) \label{eq:bc6}
\end{align}
\end{subequations}
where all the coefficients are functions of the angles. Here, $\gamma_{AB}\dd x^{A}\dd x^{B} =\dd \theta ^2+\sin ^2\theta \dd \varphi ^2$ denotes the metric on the round two-sphere, and $\gamma$ its determinant. In addition to these boundary conditions, one has to impose parity conditions on the various leading order functions of the angles in order to ensure a finite symplectic form \cite{Perez:2021abf}. This is analogous to the procedure by Regge and Teitelboim for general relativity \cite{Regge:1974zd}. They are 
\begin{align}
    &f_{rr},f_{\theta \theta },f_{\varphi \varphi },p ^{\theta \varphi },p ^{r\theta }\quad \text{(even)} &
    &f_{\theta \varphi }, p^{rr}, p^{\theta \theta }, p^{\varphi \varphi }, p^{r\varphi }\quad \text{(odd)} 
\end{align}
where even (odd) parity in terms of spherical harmonics $Y_{\ell m}$ means that $\ell$ has to be even (odd). 

The boundary condition preserving transformations in a Hamiltonian formulation are generated by  
\begin{align}
    G[\xi ,\xi ^i]=\int \dd ^3x\, \Big(\xi \mathcal{H}+\xi ^i\mathcal{H}_i\Big)+Q[\xi ,\xi ^i]
    \label{eq:whynolabel}
\end{align}
with the constraints given by \eqref{eq:constraints} and a boundary term $Q[\xi,\xi ^i]$ that will be determined shortly. For deriving the local gauge transformations of the canonical fields one may assume that both $\xi$ and $\xi^i$ fall off sufficiently fast such that all boundary terms can be dropped and compute 
\begin{align}
    \delta _{\xi }g_{ij}&= \{g_{ij},G[\xi ,\xi ^k]\}=\nabla _i \xi _j+\nabla _j \xi _i \label{eq:trafo1}\\
    \delta_{\xi}\pi^{ij}&=\{\pi^{ij},G[\xi,\,\xi^k]\}= 
    -\xi\sqrt{g}\Big(R^{ij}-\frac12g^{ij}R\Big)+\sqrt{g}\big(\nabla^i\nabla^j\xi-g^{ij}\nabla^2\xi\big)\nonumber\\
    &\qquad\qquad\qquad\qquad\;\,+\nabla_k\big(\xi^k\pi^{ij}\big) -\pi^{jk}\nabla_k\xi^i-\pi^{ik}\nabla_k\xi^j\label{eq:trafo2}\,.
\end{align}
Demanding that the above boundary and parity conditions hold yields residual transformations
\begin{align}\label{eq:res_trafo}
    \xi &=r(\Vec{\beta }\cdot \hat{r})+T+f(x^A)+\mathcal{O}(r^{-1})\\
    \xi ^r&=\Vec{\alpha }\cdot \hat{r}+W(x^A)+\mathcal{O}(r^{-1})\\
    \xi ^A&=\frac{\epsilon ^{AB}}{\sqrt{\gamma }}\partial _B(\Vec{\omega }\cdot \hat{r})+\frac{\gamma ^{AB}\partial _B(\Vec{\alpha }\cdot \hat{r})}{r}+\frac{\gamma ^{AB}\partial _B W(x^C)}{r}+\mathcal{O}(r^{-2})
\end{align}
where $\Vec{\omega }$, $\Vec{\beta }$,  $T$, $\Vec{\alpha }$ are arbitrary constants, $f(x^A)$ is a parity odd function, $W(x^A)$ is a parity even function, $\hat{r}$ denotes the unit normal of the 2-sphere, $\hat{r}=(\sin\theta \cos\varphi, \sin\theta \sin\varphi,\cos\theta)$, and $\epsilon_{AB}$ is the Levi-Civita symbol with $\epsilon_{\theta\varphi}=1$. The boundary term $Q[\xi,\xi^i]$ is fixed by demanding \eqref{eq:trafo1}, \eqref{eq:trafo2} to hold for the fall-off of the gauge parameters given by \eqref{eq:res_trafo}.

\subsection{Asymptotic symmetries}\label{sec:ASS}
The generator $G[\xi ,\xi ^i]$ should be functionally differentiable for the given boundary conditions. This condition amounts to demanding $\delta G[\xi ,\xi ^i] = \int[(...)\delta g_{ij}+(...)\delta \pi ^{ij}]$ without any boundary terms appearing. Referring to \cite{Perez:2021abf} for more details one finds that this fixes the boundary term in \eqref{eq:whynolabel} as
\begin{align}
    Q[T,\,\Vec{\alpha},\,\Vec{\beta},\,\Vec{\omega}]
    &=2\oint \dd \Omega\, \Big[Tf_{rr}+(\Vec{\omega }\cdot \hat{r})\epsilon_{AB}D^Ap^{rB}_{(-2)}+\frac{\Vec{\alpha }\cdot \hat{r}}{\sqrt{\gamma }}\Big(p^{rr}-D_Ap^{rA}\Big)\nonumber\\
    &\qquad \qquad \quad  + (\Vec{\beta} \cdot \hat{r})\Big(f_{rr}^{(-2)}+D^Af_{rA}^{(-1)}+f^A{}_A\Big)\Big] \label{eq:boundary_charge}
\end{align}
where the functions $f$ and $W$ drop out and $D_A$ is the covariant derivative associated to $\gamma_{AB}$. The set of asymptotic symmetries is then defined as the residual transformations modulo the transformations that lead to a vanishing boundary charge $Q[\xi,\xi^i]$ (the proper gauge symmetries). Non-vanishing contributions in \eqref{eq:boundary_charge} make up the physical boundary charges and allow identifying different physical states. Their algebra may be computed by evaluating the action of the residual symmetries on the fields appearing in $Q[\xi ,\xi ^i]$ and using that $\delta_\eta Q[\xi,\xi^i]=\{Q[\xi,\xi^i],Q[\eta,\eta^i]\}$. Then, under the identification 
\begin{align}
    E&=2\oint \dd ^2x \sqrt{\gamma } \, f_{rr} & J_I&=2\oint \dd ^2x \sqrt{\gamma } \, \hat{r}_I\epsilon  _{AB}D^Ap^{rB}_{(-2)} \\
    P_I&=2\oint \dd ^2x\,\hat{r}_I\big(p^{rr}-D_Ap^{rA}\big)  & B_I&=2\oint \dd ^2x \sqrt{\gamma }\,\hat{r}_I\big(f_{rr}^{(-2)}+D^Af_{rA}^{(-1)}+f^A{}_A\big)
\end{align}
we have 
\begin{align}
    Q[T,\,  \Vec{\alpha },\,\Vec{\beta },\,\Vec{\omega }]=TE+\Vec{\alpha }\cdot \Vec{P}+\Vec{\beta }\cdot \Vec{B}+\Vec{\omega }\cdot \Vec{J}
\end{align}
and one can show that the charges satisfy the finite Carroll algebra,
\begin{align}
    \{J_I,\,J_J\}&=-\epsilon_{IJK}J\,_K & \{P_I,\,B_J\}&=\delta _{IJ}E \\
    \{J_I,\,B_J\}&=-\epsilon_{IJK}\,B_K & \{J_I,\,P_J\}&=-\epsilon_{IJK}\,P_K \,.
\end{align}
Below, we provide examples for solutions of magnetic Carroll gravity where some of these charges are non-vanishing, in particular the angular momentum charges $J_I$.

\subsection{Rotating ``flat'' Carroll spacetime}\label{sec:flat_rot}
Let us consider the data
\begin{align}
    g_{ij}\dd x^i\dd x^j&=\dd r^2+r^2\dd \Omega ^2 & N&=1 & N^i&=(0,0,0) ~.
\end{align}
Since the Levi-Civita connection associated to $g_{ij}$ is Riemann flat this clearly solves the Hamiltonian constraint. Also, the constraint \eqref{eq:Keq0} is satisfied. The dynamical equation \eqref{eq:pidot} is fulfilled provided $\dot{\pi }^{ij}=0$ and what remains to solve is therefore the momentum constraints $\mH_i=0$. 

Since we are especially interested in obtaining a non-zero rotational charge let us switch on only the corresponding field in $\pi^{rA}$ and impose
\begin{align}\label{eq:rot_part}
    \pi ^{rA}=\frac{p^{rA}_{(-2)}}{r^2} ~.
\end{align}
Insertion in the constraints leads to
\begin{align}
    \mathcal{H}_A&=D_B\pi ^B{}_A =0\\
    \mathcal{H}_r&=\partial _r \pi ^{rr}-r \pi ^{AB}\gamma _{AB}+\frac{1}{r^2}D_Bp^{Br}_{(-2)}=0
\end{align}
and one can check that, e.g., 
\begin{align}
    \pi ^{AB}&=0 & \pi ^{rr}&=p^{rr}+\frac{1}{r}D_Ap^{Ar}_{(-2)}
\end{align}
with $p^{Ar}_{(-2)}(x^C)$ and $p^{rr}(x^C)$ left unfixed solves them. The non-vanishing charges for this spacetime are then given by
\begin{align}
    P_I&=2\oint \dd ^2x\,\hat{r}_I\, p^{rr} & J_I&=2\oint \dd ^2x \sqrt{\gamma } \, \hat{r}_I\epsilon _{AB}D^Ap^{rB}_{(-2)} ~.
\end{align}
If we choose additionally 
\begin{align}\label{eq:prth_choice}
    p^{rr}=0=p^{r\theta }_{(-2)} && p^{r\varphi }_{(-2)}=\frac{2J}{3\pi ^2} \sin ^2 \theta 
\end{align}
we get a single non-vanishing charge given by 
\begin{align}
    J_I|_{I=3}=J
\end{align}
describing a spacetime rotating around the $x^3$-axis.

Thus, even though the metric variables appear to describe a flat Carrollian spacetime, we have obtained a non-trivial rotational boundary charge by a suitable choice of the momenta. In fact, it can be shown that this geometry is curved once one translates the variables to covariant quantities that describe magnetic Carroll gravity in a second-order formulation. We shall provide such a description in Section \ref{sec:bulk_interp}.

\subsection{Rotating Carroll--Schwarzschild black hole}\label{sec:CSS}

Spurred by our ability to convert flat Carrollian spacetime into a rotating one, let us look at a rotating version of Carroll--Schwarzschild black holes. We pick the variables
\begin{align}\label{eq:Css_met}
    g_{ij}\dd x^i\dd x^j=\frac{\dd r^2}{f(r)}+r^2\dd \Omega ^2 && N=\sqrt{f(r)} && N^i=(0,0,0) 
\end{align}
where $f(r)=1-\frac{r_\mathrm{s}}{r}$. Since $R[g]=0$, $K_{ij}=0$ and the dynamical equation \eqref{eq:pidot} is satisfied this again solves three of the four sets of equations. We only need to solve the momentum constraints $\mathcal{H}_i=0$. As we are interested in adding a rotating charge, we choose to only switch on the momentum components \eqref{eq:rot_part} again. Then, the constraints read
\begin{align}
    \mathcal{H}_A&=D_B\pi ^B{}_A &
    \mathcal{H}_r&=\partial _r\pi^{rr}-\frac{1}{2}\pi ^{rr}\partial _r \log f(r)-rf(r)\pi ^{AB}\gamma _{AB}+\frac{1}{r^2}D_Bp^{Br}_{(-2)} ~. \label{eq:ham_r}
\end{align}

We again choose $\pi^{AB}=0$ and find the solution
\begin{align}
    \pi ^{rr}=p^{rr}+\frac{1}{r}\big(D_Bp^{Br}_{(-2)}-\tfrac12 p^{rr}r_\mathrm{s}\big)+\mathcal{O}(r^{-2}) 
\end{align}
where $p^{rr}$ and $p^{rA}_{(-2)}$ are again arbitrary functions of the angles, up to the parity conditions on $p^{rr}$. The subleading terms in $\pi ^{rr}$ are all determined by \eqref{eq:ham_r}. Since the function $p^{rr}$ is again associated to a linear momentum charge we turn it off, putting the black hole in a rest frame. For $p^{rA}_{(-2)}$ we make the choice \eqref{eq:prth_choice} and since $f_{rr}=r_\mathrm{s}$ and $f_{rr}^{(-2)}=r_\mathrm{s}^2$ we find the non-zero charges
\begin{align}
    E=8\pi r_\mathrm{s} && J_3=J
\end{align}
corresponding to a Carroll--Schwarzschild black hole rotating around the $z$-axis. This geometry has a Carroll clock form and Carroll vector field given by $v=-\frac{1}{\sqrt{f(r)}}\partial _t $ so the Carroll extremal surface at $f(r)=0$ (see \cite{Ecker:2023uwm}) is unchanged by switching on the rotational charge. The canonical momenta are given by only two non-vanishing components,
\begin{align}
    \pi ^{r\varphi }=\frac{2J}{3r^2\pi ^2}\sin ^2\theta =\pi ^{\varphi r} 
\end{align}
which are independent of $r_\mathrm{s}$.

\subsection{Geometric interpretation}
\label{sec:bulk_interp}
So far, the definition of rotation has relied on the analysis of conserved charges defined as surface integrals. One may gain a bit more intuition once one looks at the geometry in a second-order formulation. The action of magnetic Carroll gravity in this case is given by\footnote{The $(d+1)$-dimensional volume form is defined by $\mathfrak{e}:=\sqrt{\det (\tau _\mu \tau _\nu +h_{\mu \nu })}$.}
\begin{align}
    S_{\mathrm{2nd}}[h_{\mu \nu }, \tau _\mu ,C_{\mu \nu }]=\int \dd ^{d+1}x\, \mathfrak{e}\, h^{\mu \nu }\big(\mathcal{R}_{\mu \nu }[\Gamma ]-v^\rho \tau_\sigma \mathcal{R}[\Gamma ]^\sigma {}_{\mu \rho \nu }\big)
\end{align}
where $\Gamma ^\alpha {}_{\mu \nu }$ is a compatible connection\footnote{We require compatibility with the Carroll boost invariant metric data, i.e., $\nabla _\mu v^\nu =0=\nabla _\mu h_{\alpha \beta }$.} given by 
\begin{align}\label{eq:carr_conn}
    \Gamma ^\alpha {}_{\mu \nu }=&-v^\alpha \partial _{(\mu }\tau _{\nu )}-v^\alpha \tau _{(\mu }\mathcal{L}_v \tau _{\nu )}-v^\alpha C_{\mu \nu }
    +\frac{1}{2}h^{\rho \alpha }\big(\partial _\mu h_{\rho \nu }+\partial _\nu h_{\rho \mu }-\partial _\rho h_{\mu \nu }\big)-h^{\alpha \rho }K_{\rho \mu }\tau _\nu 
\end{align}
where $K_{\mu \nu }:=-\frac12 \mathcal{L}_vh_{\mu \nu }$ defines the extrinsic curvature associated to the Carroll spacetime and $\tau _\mu $ is the Carroll clock one-form. The latter satisfies orthonormality and completeness relations $\tau _\mu v^\mu =-1$, $\delta ^\nu _\mu +v^\nu \tau _\mu =h^\nu _\mu $ where $h^\nu _\mu $ is a projector onto the transverse subspace in the kernel of $v^\mu $. The symmetric tensor field $C_{\mu \nu }$ satisfies $v^\mu C_{\mu \nu }=0$ but is otherwise undetermined off-shell. Therefore, it is independent geometric data. It can be shown \cite{Bergshoeff:2017btm,Campoleoni:2022ebj,Ecker:2025vnl} that it plays the role of a Lagrange multiplier enforcing $K_{\mu \nu }=0$. By translating this action into the Hamiltonian formulation of Subsection \ref{subsec:mag} through 
\begin{align}\label{eq:translation}
    h_{\mu \nu }=\begin{pmatrix}
        N^iN_i & N_i \\ 
        N_j & g_{ij} 
    \end{pmatrix}
    && \tau =N\dd t && C_{\mu \nu }=\frac{1}{\sqrt{g}}\Big(\pi _{ij}-\frac12 g_{ij}\pi \Big)\delta ^i_\mu \delta ^j_\nu 
\end{align}
together with $v^\mu =(-1/N,N^i/N)$ it becomes evident that the part of the geometry encoding the rotational charges (the fields $\pi ^{ij}$) resides precisely in these connection degrees of freedom. Thus, a curious property of magnetic Carroll gravity is that one can dress spacetimes with different choices of connections and in this way switch on rotational charges. 

As an example, dressing the rotating Carroll--Schwarzschild solution from Subsection \ref{sec:CSS} with the symmetric tensor
\begin{align}
    C_{r\varphi}=\frac{2J\sin^3 \theta }{3r^2\pi ^2\,\sqrt{f(r)}}=C_{\varphi r}
\end{align}
yields the second-order connection
\begin{subequations}\label{eq.aff_conn_CSS}
\begin{align}
    \Gamma ^t{}_{tr}&=\frac{r_\mathrm{s}}{2r^2f(r)} & \Gamma ^t{}_{r\varphi}&=\frac{2J\sin^3 \theta }{3r^2\pi ^2\,f(r)} & & \\
    \Gamma ^r{}_{rr}&=-\frac{r_\mathrm{s}}{2r^2f(r)} & \Gamma ^r{}_{\theta \theta }&=-rf(r) & \Gamma ^r{}_{\varphi \varphi }&=-rf(r)\sin ^2\theta \\
    \Gamma ^\theta {}_{r\theta }&=\frac{1}{r} & \Gamma ^\theta {}_{\varphi \varphi }&=-\cos \theta \sin \theta & & \\
    \Gamma ^\varphi {}_{r\varphi }&=\frac{1}{r}& \Gamma ^\varphi {}_{\theta \varphi }&=\cot \theta & & 
\end{align}
\end{subequations}
with implied symmetry in the lower indices. The Riemann and Ricci tensor associated with this connection satisfy  
\begin{align}
    \mathcal{R}^\alpha {}_{\beta \mu \nu }[\Gamma ]\sim \mathcal{O}(r^{-1}) && \mathcal{R}_{\mu \nu }[\Gamma ]=0
\end{align}
such that the spacetime is asymptotically isometric to the homogeneous spacetime associated to the Carroll algebra. To our knowledge, there is no scalar curvature invariant that captures the rotational charge $J$. All quantities found in \cite{Figueroa-OFarrill:2022mcy} either vanish or are only proportional to $r_\mathrm{s}$. An exception is the analog of the Chern--Pontryagin density
\begin{align}
\mathcal{CP}=\mathfrak{e}\,\epsilon_{\mu \nu \alpha \beta}\,\mathcal{R}^{\mu \nu }{}_{\sigma \rho }\,\mathcal{R}^{\alpha \beta }{}_{\gamma \tau }\,h^{\sigma \gamma }\,h^{\rho \tau }=\frac{48Jr_\mathrm{s}\sin(2\theta)}{\pi^2r^7}
 \end{align}
 which, however, is not a Carroll boost invariant quantity. Indeed, while the local Carroll boosts are fixed in the Hamiltonian approach (because we always choose $\tau =N\dd t$) they are reappearing in the second-order formulation and all physical quantities should be invariant under them. On-shell, the connection $\Gamma ^\alpha {}_{\mu \nu } $ is torsion-free and boost-invariant (see, e.g., \cite{Ecker:2025vnl}).

\subsubsection{Geodesics}
Since rotational charges affect only the connection in the bulk it becomes natural to ask about the behavior of geodesics. Consider a curve $x^\mu (s)$ with parameter $s$ and assume, for simplicity, that the spacetime is given by the rotating Carroll--Schwarzschild solution from Subsection \ref{sec:CSS} with the connection given by \eqref{eq.aff_conn_CSS}. Restricting to the equatorial plane, $\theta =\pi/2$, the autoparallel equation 
\begin{align}
    \frac{\mathrm{d}^2 x^\mu }{\mathrm{d}s^2}+\Gamma ^\mu{}_{\nu \alpha }\,\frac{\mathrm{d} x^\nu }{\mathrm{d}s}\frac{\mathrm{d} x^\alpha }{\mathrm{d}s}=0
\end{align}
may be simplified by identifying three constants of motion $\{F,\ell ,\mE\}$. For this simple scenario, the autoparallel equation is essentially equivalent to the Carroll geodesic equations presented in \cite{Ciambelli:2023tzb} and we get
\begin{subequations}
    \label{eq:angelinajolie}
    \begin{align}
\dot{t} &= \frac{F}{1-\frac{r_\mathrm{s}}{r}} + \frac{4J\ell}{9\pi^2r^2(r-r_\mathrm{s})}  \label{eq:whatever} \\
\dot{\varphi} &= \frac{\ell}{r^2}\\
\frac{\dot r^2}{2}+V_{\textrm{\tiny eff}}(r)&=\mE
    \end{align}
with the effective potential
\begin{align}
V_{\textrm{\tiny eff}}(r)=\frac{r_\mathrm{s} \mE}{r} + \frac{\ell^2}{2r^2} - \frac{r_\mathrm{s}\ell^2}{2r^3}
\label{eq:lalapetz}
\end{align}
\end{subequations}
which coincides with Eq.~(17) in \cite{Ciambelli:2023tzb}.

Thus, the discussion of Carroll--Schwarzschild geodesics in \cite{Ciambelli:2023tzb} applies also to rotating Carroll--Schwarzschild. In particular, there are no circular orbits except on the Carroll extremal surface, which is a key difference to geodesics in Schwarzschild or Kerr backgrounds. Notably, the effective potential \eqref{eq:lalapetz} depends on the radial integration constant $\mE$; this is a small but significant difference to the effective potential generated by the Schwarzschild black hole and ultimately responsible for the absence of stable circular orbits on rotating Carroll--Schwarzschild black hole backgrounds.

The only difference to non-rotating Carroll--Schwarzschild is the way energy $F$ relates to $\dot t$. Without rotation, the last term in \eqref{eq:whatever} vanishes. With rotation, there is a critical tuning between particle energy $F$, particle angular momentum $\ell$, black hole radius $r_\mathrm{s}$, and black hole angular momentum $J$ such that the pole at the Carroll extremal surface cancels:
\begin{align}
J_{\textrm{\tiny crit}} = - \frac{9\pi^2 r_\mathrm{s}^3 F}{4\ell}
    \label{eq:Jcrit}
\end{align}
For the critical tuning \eqref{eq:Jcrit} we find that $\dot t$ has no pole at $r=r_\mathrm{s}$ but instead takes the value $\dot t=3F$ at the Carroll extremal surface. 

The last term in \eqref{eq:whatever} mirrors the corresponding term for equatorial geodesics on a Kerr black hole. In that case, however, the equation for $\dot{\varphi}$ also contains additional terms proportional to $J$ that cause frame-dragging. The absence of these terms in the present case make frame-dragging impossible for rotating Carroll--Schwarzschild black holes. Thus, we see again small but significant differences between geodesics in Carrollian black hole backgrounds and geodesics in Lorentzian counterparts.

\subsubsection{Killing fields}
Since on-shell the boost-invariant Carroll data consists of $v^\mu $, $h_{\mu \nu }$ and the connection $\Gamma ^\alpha {}_{\mu \nu }$, a Carroll Killing field should leave them invariant,
\begin{align}
    \mathcal{L}_\zeta v^\mu =0 && \mathcal{L}_\zeta h_{\mu \nu }=0 && (\mathcal{L}_\zeta \nabla )_\mu =0
\end{align}
where the last equality, called the affine Killing condition \cite{Morand:2018tke,Bekaert:2015xua}, reads explicitly 
\begin{align}
    \zeta ^\alpha \partial _\alpha \Gamma ^\sigma {}_{\mu \nu }+\Gamma ^\sigma {}_{\alpha \nu }\partial _\mu \zeta ^\alpha +\Gamma ^\sigma {}_{\mu \alpha }\partial _\nu \zeta ^\alpha -\Gamma ^\alpha {}_{\mu \nu }\partial _\alpha \zeta ^\sigma +\partial _\mu \partial _\nu \zeta ^\sigma =0 ~. 
\end{align}
From the first two we find that any $\zeta =f(r,\theta ,\varphi)\partial _t +\partial _\varphi $ is a symmetry but the last one eliminates all but two Killing vectors 
\begin{align}
    \zeta =a\partial _t + b\partial _\varphi && a,b\in \mathbb{R}
\end{align}
which is the same set as for the Kerr solution.

We can relate the diffeomorphism generated by any vector field to a surface deformation in ADM variables by taking appropriate projections, 
\begin{align}
    \xi &= \tau _\mu \zeta ^\mu =N\zeta ^t & \xi ^i&=h^i_\mu \zeta ^\mu =\zeta ^i+N^i\zeta ^t ~.
\end{align}
It is then straightforward to show that for the case of the rotating Carroll--Schwarzschild geometry
\begin{align}
    Q[\partial _\varphi ]=J && Q[\partial _t]=8\pi r_\mathrm{s} ~.
\end{align}
Thus, the presence of non-vanishing Hamiltonian boundary charges corresponds to the presence of bulk isometries, as anticipated on general grounds.  


\section{Odd-power expansion of general relativity}\label{sec:4}
In contrast to the previous Section, one may think about rotation of Carroll spacetimes in a different way. Instead of considering magnetic Carroll gravity, one may introduce an extension of it that allows describing the Carroll limit of the Kerr geometry. Our proposal for this theory stems from an odd-power-expansion in the speed of light $c$, in contrast to the even-power expansion of \cite{Hansen:2021fxi}. This will allow us to obtain a Carrollian theory that contains magnetic Carroll gravity as a subsector. We start again from the Hamiltonian form of Einstein gravity,
\begin{align}
    I = \int \dd t\, \dd^d x \big(\pi^{ij} \dot g_{ij} - N \mH  - N^i \mH_i\big)
\end{align}
with
\begin{align}\label{eq:oh}
    \mH
    = \frac{16\pi G_\mathrm{E}}{\sqrt{g}}
      \bigg(\pi^{ij} \pi_{ij} - \frac{\pi^2}{d-1}\bigg)
      - \frac{\sqrt{g} c^2}{16\pi G_\mathrm{E}}R[g] &&  \mH_i
      &= -2 g_{ik} \nabla _j \pi^{kj}\, .
\end{align}
Here, $d$ is the number of spatial dimensions, $R$ is the Ricci scalar of $g_{ij}$, and we expand around the electric limit of Carroll gravity. Hence, the gravitational coupling constant is the rescaled ``electric'' version $G_\mathrm{E}$ instead of $G_\mathrm{M}$.

Following the logic of the  pre-ultra-local (PUL) parametrization in \cite{Hansen:2021fxi}, with the modification that we allow for odd powers of $c$ as in the non-relativistic expansion of  \cite{Hansen:2020wqw}, we expand the spatial metric $g_{ij}$, its canonical momentum $\pi^{ij}$, as well as the lapse $N$ and shift $N^i$ in odd powers of $c$,
\begin{subequations}\label{eq:odd-power-expansion-of-fields}
    \begin{align}
    g_{ij} &= \glo_{ij} + c\,m_{ij} + c^2 l_{ij} + \mO(c^3) \\
    \pi^{ij} &= \pilo^{ij} + c\,\rho^{ij} + c^2 \sigma^{ij} + \mO(c^3) \\
    N &= N_{(0)} + c\,N_{(1)} + c^2 N_{(2)} + \mO(c^3) \\
    N^i &= N_{(0)}^i + c\,N_{(1)}^i + c^2 N_{(2)}^i + \mO(c^3)\,.
\end{align}
\end{subequations}
All spatial indices are raised and lowered using the leading order metric $\glo_{ij}$ and its inverse $\glo^{ij}$. This fixes the expansion of the inverse metric $g^{ij}$ through the requirement $g_{ik} g^{kj} = \delta^j_i$ as
\begin{align}
    g^{ij}
    = \glo^{ij}
      - c\,\glo^{ik} \glo^{jl} m_{kl}
      + c^2\big(\glo^{ik} \glo^{jp} \glo^{ln}m_{kl}m_{np} - \glo^{ik} \glo^{jl} l_{kl}\big)
      + \mO(c^3) \,.
\end{align}
It is convenient to also expand the DeWitt supermetric $G_{ijkl}$ in powers of $c$,
\begin{align}
    G_{ijkl} = H_{ijkl} + c\,M_{ijkl} + c^2 L_{ijkl} + \mO(c^3)
\end{align}
where
\begin{align}
    H_{ijkl} &= \frac{1}{2}\Big(
        \glo_{ik} \glo_{jl}
      + \glo_{il} \glo_{jk}
      - \frac{2}{d-1} \glo_{ij} \glo_{kl}
    \Big) \\
    M_{ijkl} &= \frac{1}{2}\Big(
        \glo_{ik} m_{jl} + \glo_{jl} m_{ik}
      + \glo_{il} m_{jk} + \glo_{jk} m_{il}
      - \frac{2}{d-1}\left(\glo_{ij} m_{kl} + \glo_{kl} m_{ij}\right)
    \Big) \\
    L_{ijkl} &= \frac{1}{2}\Big(
        \glo_{ik} l_{jl} + \glo_{jl} l_{ik}
      + m_{ik} m_{jl}
      + \glo_{il} l_{jk} + \glo_{jk} l_{il}
      + m_{il} m_{jk}
      \nonumber\\
      & \qquad \qquad \qquad \qquad \qquad \qquad - \frac{2}{d-1}\left(
          \glo_{ij} l_{kl} + \glo_{kl} l_{ij} + m_{ij} m_{kl}
        \right) \Big) 
\end{align}
such that $H_{ijkl}$ is the usual DeWitt supermetric built from $\glo_{ij}$ and $M_{ijkl},L_{ijkl}$ encode the dependence on $m_{ij}$ and $l_{ij}$ at subleading orders. We also define the scalar coefficients
\begin{align}
    A &= -\frac{1}{2} \glo^{ij} m_{ji} &
    B &= \frac{1}{8}\big(\glo^{ij} m_{ji}\big)^2
       + \frac{1}{4} \glo^{ij} m_{jk} \glo^{kl} m_{li}
       - \frac{1}{2} \glo^{ij} l_{ji}
\end{align}
which arise from the expansion of $\sqrt{g} = \sqrt{\glo}\,(1-cA-c^2(B-A^2)+\mathcal O(c^3))$ and $g^{ij}$ in powers of $c$ for notational simplicity. Expanding the symplectic form $\pi^{ij} \dot g_{ij}$, we find
\begin{align}
    \pi^{ij} \dot g_{ij}
    &= \pilo^{ij} \dot \glo_{ij}
     + c\big(\pilo^{ij} \dot m_{ij} + \rho^{ij} \dot \glo_{ij}\big)
     + c^2\big(\pilo^{ij} \dot l_{ij} + \rho^{ij} \dot m_{ij} + \sigma^{ij} \dot \glo_{ij}\big)
     + \mO(c^3)\,.
\end{align}
At order $c^2$, we use this to identify three canonical pairs, leading to the fundamental non-vanishing Poisson brackets
\begin{subequations}
\begin{align}
    \{\glo_{ij}(x), \sigma^{kl}(y)\}
      &= \delta^k_{(i} \delta^l_{j)} \,\delta(x-y) \\
    \{m_{ij}(x), \rho^{kl}(y)\}
      &= \delta^k_{(i} \delta^l_{j)} \,\delta(x-y) \\
    \{l_{ij}(x), \pilo^{kl}(y)\}
      &= \delta^k_{(i} \delta^l_{j)} \,\delta(x-y)\,.
\end{align}
\end{subequations}
All other brackets vanish. The lower orders give rise to different canonical pairs.  Thus, $(\glo_{ij},\sigma^{ij})$, $(m_{ij},\rho^{ij})$ and $(l_{ij},\pilo^{ij})$ form three copies of the usual ADM phase space. 
The first class constraints expand as
\begin{align}
    \mH &= \mH^{(0)} + c\,\mH^{(1)} + c^2 \mH^{(2)} + \mO(c^3) &
    \mH _i&= \mH_i^{(0)} + c\,\mH_i^{(1)} + c^2 \mH_i^{(2)} + \mO(c^3) ~.
\end{align}
Explicitly, the scalar constraints are
\begin{subequations}
\label{eq:NNLO-gravity-energy-constraints}
\begin{align}
    \mH^{(0)}
      &= \mH_\mathrm{E}
      = \frac{16\pi G_\mathrm{E}}{\sqrt{\glo}}\,H_{ijkl} \pilo^{ij}\pilo^{kl} \label{eq:aha}\\
    \mH ^{(1)}
      &= \frac{16\pi G_\mathrm{E}}{\sqrt{\glo}}\Big(
          H_{ijkl}\big(A \pilo^{ij}\pilo^{kl} + 2 \pilo^{ij}\rho^{kl}\big)
        + M_{ijkl} \pilo^{ij}\pilo^{kl}
        \Big)\\
    \mH^{(2)}
      &= \frac{16\pi G_\mathrm{E}}{\sqrt{\glo}}\Big(
          H_{ijkl}\big(
              2\pilo^{ij}\sigma^{kl}
            + \rho^{ij}\rho^{kl}
            + 2A \pilo^{ij}\rho^{kl}
            + B \pilo^{ij}\pilo^{kl}
          \big) \nonumber
         \\ &\quad +\,  M_{ijkl}\big(
              2\pilo^{ij}\rho^{kl}
            + A \pilo^{ij}\pilo^{kl}
          \big)
        + L_{ijkl}\pilo^{ij}\pilo^{kl}
        \Big) 
      - \frac{\sqrt{\glo}}{16\pi G_\mathrm{E}}R[\glo] \label{eq:oho}
\end{align}
\end{subequations}
where $R[\glo]$ is the Ricci scalar of $\glo_{ij}$. The various orders of the momentum constraints read
\begin{subequations}
\label{eq:NNLO-gravity-momentum-constraints}
\begin{align}
    \mH_k^{(0)}
      &= -2 \glo_{ik} \nabla_j \pilo^{ij} \\[0.3em]
    \mH_k^{(1)}
      &= -2 \glo_{ik} \nabla_j \rho^{ij}
         - 2 m_{ik} \nabla_j \pilo^{ij}
         + C^{(m)}_{ijk} \pilo^{ij} \\[0.3em]
    \mH_k^{(2)}
      &= -2 \glo_{ik} \nabla_j \sigma^{ij}
         - 2 m_{ik} \nabla_j \rho^{ij}
         - 2 l_{ik} \nabla_j \pilo^{ij}
         + C^{(m)}_{ijk} \rho^{ij}
         + C^{(l)}_{ijk} \pilo^{ij}
\end{align}
and contain the covariant derivative with respect to $\glo_{ij}$, denoted by $\nabla_i$ as well as the definition
\begin{align}
    C_{ijk}^{(q)} = \nabla_k q_{ij} - \nabla_j q_{ik} - \nabla_i q_{jk}
\end{align}
\end{subequations}
for $q_{ij}  = \{m_{ij}, l_{ij}\}$. It has been shown in the second order formalism that the higher order action in such expansions systematically reproduces the equations of motion from the lower order actions. Hence, by studying the dynamics of e.g. the order-$c^2$-action, one automatically obtains the lower order dynamics as well \cite{Hansen:2020pqs}. The complete set of Poisson brackets between these first class constraints is given in Appendix~\ref{sec:Poisson-brackets}. One may also expand the generator of gauge transformations\footnote{The variable $\xi$ represents collectively the components $(\xi ,\xi^i)$ defined in \eqref{eq:whynolabel}.} by 
\begin{align}
    G[\xi ]=G^{(0)}[\xi _{(0)}] + c\, G^{(1)}[\xi _{(0)}, \xi _{(1)} ] + c^2\, G^{(2)}[\xi _{(0)},\xi _{(1)},\xi _{(2)}]
\end{align}
where we assumed a similar expansion of the gauge parameters $\xi =\xi _{(0)}+c\, \xi _{(1)}+c^2\, \xi _{(2)}$. In particular, the generator of spatial diffeomorphisms reads at $\mathcal{O}(c^2)$
\begin{align}\label{eq:diffgen}
    G^{(2)}[\xi _{(0)},\xi _{(1)},\xi _{(2)}]=\int \dd ^d x\, \Big(\xi ^i_{(0)} \mH _i^{(2)} + \xi ^i _{(1)}\mH _i^{(1)}+\xi ^i_{(2)}\mH _i^{(0)}\Big)
\end{align}
showing that within the $\mathcal{O}(c^2)$-theory $\mH _i^{(2)}$ generates leading order diffeomorphisms while $\mH_i^{(1)}$ is associated to NLO diffeomorphisms. Typically (see e.g. Section 2.2 in \cite{Hansen:2020pqs} for the nonrelativistic gravity context) one regards all the subleading transformations as additional gauge transformations and only interprets the leading order as being associated to the diffeomorphism freedom of the expanded theory. This is consistent with the action of the respective generator $D^{(2)}[\xi ]:=\int \dd ^dx\, \xi ^i\mH_i ^{(2)}$ on the canonical fields, $\{\Phi (x),D^{(2)}[\xi ]\}=\mathcal{L}_\xi \Phi (x)$, as shown in Appendix \ref{sec:Diff-generators}. 

Since we have thrice the number of field variables and momenta, $(\glo_{ij}, m_{ij}, l_{ij})$ and $(\pilo^{ij},\rho^{ij},\sigma^{ij})$, and also thrice the number of first class constraints, the number of local physical degrees of freedom has also tripled compared to magnetic Carroll gravity 
\begin{align}
    \mathrm{d.o.f.}_\text{NNLO}
    = \frac{3}{2} D(D-3), 
\end{align}
where $D=d+1$.
Before turning to the next Subsection a few remarks are in order. Looking at the expansion of the Hamiltonian constraint it is evident that the leading order term \eqref{eq:aha} reproduces the electric Hamiltonian $\mH_E$, cf. \eqref{eq:electricH}. Moreover, part of the NNLO constraint $\mH ^{(2)}$ in \eqref{eq:oho} reproduces the magnetic Hamiltonian $\mH _M$ of \eqref{eq:constraints}. This hierarchical structure is characteristic of the Carroll expansion of general relativity and has previously been observed in a second order formulation \cite{Hansen:2021fxi}. In particular, magnetic Carroll gravity can be obtained equivalently as a consistent truncation of the $\mathcal{O}(c^2)$ theory when restricting the expansion to even powers of $c$ and excising the electric sector. What is new in the present analysis is the emergence of an intermediate order that has not been captured previously. In Subsection \ref{subsec:4.1} we shall mimic the truncation done for arriving at the magnetic theory but take the fields at this intermediate order into account as well. As we shall see, this gives rise to an extension of magnetic Carroll gravity that may capture rotational properties. 

\newpage

\subsection{Extended magnetic Carroll gravity}\label{subsec:4.1}
In the following, we shall restrict to a subsector of the NNLO theory obtained by imposing the constraint
\begin{align}\label{eq:weak-magnetic-condition}
    \pilo^{ij} = 0 
\end{align}
on the leading order canonical momentum. By comparing with \eqref{eq:aha} this truncates away the electric sector. 
At the level of the Hamiltonian, it implies
\begin{align}\label{eq:Postmagnetic-LO-NLO}
    \mH^{(0)} &= \mH_i^{(0)} =  
    \mH^{(1)} = 0 & \mH^{(1)}_{i} &= -2  \glo_{ik}\nabla_j\rho^{jk}
\end{align}
such that the action up to $\mO(c^2)$ reads
\begin{align}
    I&=c\int \dd t\; \dd^d x \,\Big[\rho ^{ij}\Dot{h}_{ij}+2N^i_{(0)} \glo_{ik} \nabla_j \rho^{kj} \Big] 
    +c^2 \int \dd t\; \dd^d x \,\mathcal{L}_{(2)}
    + \mO(c^3) ~.
\end{align}
This contains the $\mO(c^2)$-Lagrangian density
\begin{align}
    \mathcal{L}_{(2)}
    &= \rho^{ij} \dot m_{ij} + \sigma^{ij}\dot{h}_{ij} \nonumber \\
    &-N_{(0)}\bigg(
         \frac{16\pi G_\mathrm{E}}{\sqrt{\glo}} H_{ijkl}\rho^{ij}\rho^{kl}
         - \frac{\sqrt{\glo}}{16\pi G_\mathrm{E}}R
       \bigg)
       + 2N^i_{(1)} \glo_{ik}\nabla_j \rho^{jk} \nonumber\\
    &\quad
       + 2N^i_{(0)}\big(
           \glo_{ik}\nabla_j \sigma^{jk}
           + m_{ik}\nabla_j\rho^{jk}
           - \frac{1}{2} C^{(m)}_{kji}\rho^{kj}
       \big)
\end{align}
which defines the action of extended magnetic Carroll gravity, 
\begin{align}\label{eq:exMaction}
    I_\mathrm{exM}[\rho ^{ij}, m_{ij}, \sigma ^{ij}, h_{ij}, N_{(0)}, N_{(1)}^i, N_{(0)}^i]:=\int \dd t \dd ^d x\, \mathcal{L}_{(2)} ~.
\end{align} 
This action contains both scalar and momentum constraint contributions. The Hamiltonian constraint reads
\begin{align}\label{eq:extmag-Hamiltonian-constraint}
    \mH^{(2)}
    = \frac{16\pi G_\mathrm{E}}{\sqrt{\glo}} H_{ijkl}\rho^{ij}\rho^{kl}
      - \frac{\sqrt{\glo}}{16\pi G_\mathrm{E}}R[h]
\end{align}
and can be seen to reduce to the standard magnetic Carroll gravity Hamiltonian when setting the odd power momentum field $\rho ^{ij}$ to zero. The same is true for the associated momentum constraints. However, note that the ``electric'' coupling constant $G_\mathrm{E}$ features in \eqref{eq:extmag-Hamiltonian-constraint}, not the magnetic counterpart $G_\mathrm{M}$.
The full set of constraints featuring in $I_{\mathrm{exM}}$ closes under Poisson brackets and realizes an algebra reminiscent of the Carrollian surface deformation algebra.
In particular, the NNLO Hamiltonian constraint commutes with itself
\begin{subequations}
\begin{align}\label{eq:Carroll-condition}
    \{\mH^{(2)}(x),\,\mH^{(2)}(y)\} &= 0\,.
\end{align}
For the remaining Poisson brackets one obtains
\begin{align}
  \{\mH^{(2)}(x),\,\mH^{(2)}_i(y)\} &= \mH^{(2)}(y)\,\partial_i \delta(x-y)\\
  \{\mH^{(2)}_i(x),\,\mH^{(2)}_j(y)\}
    &= \mH^{(2)}_i(y)\,\partial_j \delta(x-y)
     + \mH^{(2)}_j(x)\,\partial_i \delta(x-y)\\
  \{\mH^{(2)}_i(x),\,\mH^{(1)}_j(y)\}
    &= \mH^{(1)}_i(y)\,\partial_j \delta(x-y)
     + \mH^{(1)}_j(x)\,\partial_i \delta(x-y)\\
  \{\mH^{(2)}(x),\,\mH^{(1)}_i(y)\} &= 0\\
  \{\mH^{(1)}_i(x),\,\mH^{(1)}_j(y)\} &= 0\,.
\end{align}
\end{subequations}
This reduces to the magnetic Carroll algebra in the limit of vanishing odd-power fields \cite{Henneaux:2021yzg,Campoleoni:2022ebj, Perez:2021abf}. By means of their closure under the Poisson bracket the constraints $\mH_i^{(2)}$ may be thought of as generating spatial diffeomorphisms, as expected from the discussion around \eqref{eq:diffgen}.
The equations of motion derived from $I_{\mathrm{exM}}$, 
\begin{subequations}
\begin{align}
    \dot \glo_{ij}
    &=
    \mathcal L_{N_{(0)}}\glo_{ij}
    \\
    \dot m_{ij}
    &=
    \frac{32\pi G_\mathrm{E}}{\sqrt h}\,
    N_{(0)}
    \big(
        \rho_{ij}
        -\frac{1}{d-1}\glo_{ij}\rho
    \big)
    +
    \mathcal L_{N_{(1)}}\glo_{ij}
    +
    \mathcal L_{N_{(0)}}m_{ij}
    \\
    \dot\rho^{ij}
    &=
    \mathcal L_{N_{(0)}}\rho^{ij}
    \\
    \dot\sigma^{ij}
    &=
    \frac{\sqrt h}{16\pi G_\mathrm{E}}
    \Big[
        -N_{(0)}
        \Big(
            R^{ij}
            -\frac12 \glo^{ij}R
        \Big)
        +
        \nabla^i\nabla^jN_{(0)}
        -
        \glo^{ij}\nabla^2N_{(0)}
    \Big]
    \nonumber\\
    &\quad
    -
    \frac{16\pi G_\mathrm{E} N_{(0)}}{\sqrt h}
    \Big[
        2\rho^i{}_{k}\rho^{jk}
        -
        \frac{2}{d-1}\rho\,\rho^{ij}
        -
        \frac12 \glo^{ij}
        \Big(
            \rho^{kl}\rho_{kl}
            -
            \frac{\rho^2}{d-1}
        \Big)
    \Big]
    \nonumber\\
    &\quad
    +
    \mathcal L_{N_{(1)}}\rho^{ij}
    +
    \mathcal L_{N_{(0)}}\sigma^{ij}
\end{align}
\end{subequations}
are of first-order in time derivatives by construction. Here $\rho^{ij}$ and $\sigma^{ij}$ are again tensor densities of weight $+1$ and we defined $\rho =\glo^{ij}\rho_{ij}$.

The first equation implements the constraint that the leading order extrinsic curvature vanishes, $K_{ij}=0$, which survives this extension of magnetic Carroll gravity. The remaining equations govern the dynamics of the additional odd-power fields $(m_{ij},\rho^{ij},\sigma^{ij})$. The degrees of freedom of extended magnetic Carroll gravity can now be counted as follows. The canonical variables consist of two symmetric tensors $\glo_{ij}$ and $m_{ij}$ and their conjugate momenta $\sigma^{ij}$ and $\rho^{ij}$, giving $2D(D-1)$ phase space variables, where $D=d+1$ denotes the number of spacetime dimensions. The full set of constraints comprises one scalar constraint $\mH^{(2)}$ and two sets of $d$ momentum constraints $\mH^{(1)}_i$ and $\mH^{(2)}_i$, i.e., a total of $2D-1$ first-class constraints, each removing two degrees of freedom in the phase space. Hence
\begin{align}
    \mathrm{d.o.f.}
    = \frac{1}{2}\big(2D(D-1) - 2(2D-1)\big)
    = D(D-3)+1.
\end{align}
Evaluating this for $D=3,\,4,\,5$ spacetime dimensions yields
\begin{align}
    \mathrm{d.o.f.}\big|_{D=3} &= 1 & \mathrm{d.o.f.}\big|_{D=4} &= 5 & \mathrm{d.o.f.}\big|_{D=5} &= 11\,.
\end{align}

As expected, the introduction of the additional field $m_{ij}$ and its momentum $\rho^{ij}$ increases the number of propagating modes compared to magnetic Carroll gravity. In $D=4$ the number of degrees of freedom curiously coincides with the number of massive graviton polarizations. In $D=3$ the theory acquires a single local degree of freedom.

\subsection{Kerroll black hole}\label{sec:Kerroll_BHs}
We now show that a $c\to 0$ expansion of the Kerr black hole leads to a solution of extended magnetic Carroll gravity, which we refer to as the \textit{Kerroll black hole}. Let us start with the Kerr metric \eqref{eq:kerr_mag_metric} in ADM variables,
\begin{align}\label{eq:3+1-decomposition-of-metric}
    \dd s^2=-N^2c^2\dd t^2+g_{ij}(\dd x^i+N^i\dd t)(\dd x^j+N^j\dd t) ~.
\end{align}
We work in Boyer-Lindquist coordinates $(t,r,\theta,\varphi)$ and again use the definitions $\Delta = r^2 - r_\mathrm{s} r + a^2$ and $\Sigma = r^2 + a^2 \cos^2\theta $ where $r_\mathrm{s}=2G_\mathrm{N}M/c^2$.  The spatial metric $g_{ij}$ is given by
\begin{align}
    g_{ij}\dd x^i \dd x^j
    = \frac{\Sigma}{\Delta}\,\dd r^2
      + \Sigma\,\dd\theta^2
      + \sin^2\theta\, \bigg(
          r^2 + a^2 + \frac{r_\mathrm{s} r a^2 \sin^2\theta}{\Sigma}
        \bigg)\,\dd\varphi^2
\end{align}
while lapse and shift are
\begin{align}
        N= \sqrt{\frac{\Sigma\Delta}{(r^2+a^2)\Sigma+r_\mathrm{s} r a^2\sin^2\theta}}\qquad\qquad  N^i = -\frac{c\,r_\mathrm{s} r a}{(r^2+a^2)\Sigma+r_\mathrm{s} r a^2\sin^2\theta}\,\delta^i_\varphi\,.
\end{align}
Finally, the ADM equations of motion determine the canonical momentum $\pi_{ij}$. It may be expressed in terms of the extrinsic curvature $\mK_{ij}$ as $\pi_{ij}=\frac{\sqrt{g}}{16\pi G_\mathrm{E}}(\mK_{ij}-g_{ij}\mK)$, where
\begin{align}
    \mK_{ij}=\frac{1}{2N}\big(\dot g_{ij}-\nabla_iN_j -\nabla_j N_i\big)\,.
\end{align}
Consistently with the expansion set up in \eqref{eq:oh}, we hold the electric combination $G_\mathrm{E}=G_\mathrm{N}c^{-2}$ fixed such that $r_\mathrm{s}=2G_\mathrm{E}M$. The parameters $M$ and $a$ are assumed to not scale with $c$.
Using the expansions of the various fields \eqref{eq:odd-power-expansion-of-fields} we then find 
\begin{subequations}\label{eq:Kerroll_sol}
    \begin{align}
        \glo_{ij}
    &= g_{ij} & m_{ij}&=0 & l_{ij}&=0 \\
        \pilo^{ij}
    &= 0 & \rho^{ij}&=\frac{\sqrt{\glo}}{16\pi G_\mathrm{E}}\,\big(S^{ij} - \glo^{ij} S\big) & \sigma^{ij}&=0 \\
    N_{(0)}&=N & N_{(1)}&=0 & N_{(2)}&=0 \\
    N^i_{(0)}&=0 & N^i_{(1)}&=-\frac{r_\mathrm{s} r a}{(r^2+a^2)\Sigma+r_\mathrm{s} r a^2\sin^2\theta}\delta^i_\varphi & N^i_{(2)}&=0
    \end{align}
\end{subequations}
where $S_{ij}$ is a shorthand for the NLO term in the expansion of $\mK_{ij}=K_{ij}+cS_{ij}+\mO(c^2)$,
\begin{align}
  S_{ij}=-\frac{1}{2N_{(0)}}\,\big(\nabla_iN_{(1)j}+\nabla_jN_{(1)i}\big)~.
\end{align} 
We therefore obtain the Carroll metric data directly from the expansion of the relativistic Kerr black hole metric data. When expanding lapse and shift, only the leading order lapse contributes, whereas the shift is again of order $c$. Crucially, the leading order canonical momentum and hence the leading order extrinsic curvature vanish with nontrivial contributions only of $\mO(c)$. This ensures compatibility with the constraint $\pilo ^{ij}=0$ imposed to define extended magnetic Carroll gravity. One can show that all the constraints as well as the evolution equations of the theory \eqref{eq:exMaction} are satisfied. The fields \eqref{eq:Kerroll_sol} thus define a two-parameter family of solutions to extended magnetic Carroll gravity.

In the context of Carroll geometry it is often convenient to work with the second order PUL parametrization as a starting point for the Carroll expansion \cite{Hansen:2021fxi}. The corresponding PUL vector field can then be expressed in terms of the lapse and shift as
\begin{align}
    V^\mu=\frac{1}{N}\,\big(-1,N^i\big)\,.
\end{align}
From Eq.~\eqref{eq:Kerroll_sol} one concludes that expanding $V^\mu$ of the Kerr black hole in powers of $c$ not only yields the Carroll vector field $v$, but also a non-zero spatial subleading field $u^i = N^i_{(1)}/N_{(0)}$, stemming from the expanded term linear in $c$. This field encodes the rotational features of the Kerroll black hole as an odd-power effect in $c$. 

The metric formalism also results in higher order contributions to the PUL spatial metric $\Pi_{\mu\nu}=g_{\mu\nu}+c^2T_\mu T_\nu$, where, after fixing the Carroll clock one-form $T_\mu=\tau_\mu$, one similarly obtains non-vanishing subleading fields $m_{\mu\nu}$ and $l_{\mu\nu}$ in the expansion $\Pi_{\mu\nu}=h_{\mu\nu}+c\,m_{\mu\nu}+c^2\, l_{\mu\nu}+\mO(c^3)$, which is also evident from Eq.~\eqref{eq:3+1-decomposition-of-metric}.  See Appendix \ref{sec:appendix-2nd-order} for more details on the odd-power expansion in the second order formalism. 

Representing the Kerroll black hole in the Carrollian ADM-like variables is more natural than in second order language, since one does not obtain multiple contributions in different orders of $c$ from the same relativistic parent field, as can be seen in Eq. \eqref{eq:Kerroll_sol}. The significance of including odd-powers in the expansion explains the difficulties in constructing the Carrollian analog of rotating Lorentzian black holes in standard magnetic Carroll gravity \cite{Ecker:2023uwm, Kolar:2025ebv}, which does not include contributions from odd-powers in $c$. 

The Carroll extremal surface
\begin{align}
    r_\mathrm{c} = \frac{r_\mathrm{s}}{2} + \sqrt{\frac{r_\mathrm{s}^2}{4}-a^2}
\end{align} 
coincides with the event horizon of the relativistic Kerr black hole. This is analogous to the non-rotating case \cite{Ecker:2023uwm}. We define a Carroll extremal surface like in the Introduction, as a locus where the Carroll vector field $v$ diverges, which corresponds to solving $\Delta =0$ here. For a more precise definition in 2d, see \cite{Ecker:2023uwm}.

\subsection{Boundary charges}
\label{subsec:bdy-charges}
One can determine the physical charges of the Kerroll black hole by examining which boundary terms are required to make the canonical generators associated with bulk symmetries functionally differentiable. Although we do not analyze the full space of asymptotic symmetries here, we can restrict attention to a minimal set that is certainly contained within it --- namely, the vector fields $\partial_t $ and $-\partial_\varphi$.

Let us start by considering the canonical generator of a diffeomorphism along $\partial_\varphi$. From the discussion around Eq.~\eqref{eq:diffgen} we know that diffeomorphisms in this theory should be generated by $\mH _i^{(2)}$, so we investigate whether $-\partial _\varphi $ is associated to a non-vanishing boundary charge of this generator. Indeed, keeping all boundary terms we may write the variation of $D^{(2)}[\xi ]$ as
\begin{align}
    \delta D^{(2)}[\xi ]&=\int \dd ^dx\, \xi ^i\delta \mH_i^{(2)}=\int \dd ^dx\, \big[(...)\delta h_{ij}+(...)\delta \rho ^{ij}+ \partial _iB^i \big] 
\end{align}
where we used that $\delta m_{ij}=0$ and $\delta \sigma ^{ij}=0$ for the family of solutions we are looking at. The total derivative term reads
\begin{align}
    B^i=-2\xi^{(j}\sigma ^{k)i}\delta h_{jk}-2\xi ^{(j}m_{jk}\delta \rho ^{k)i}+\xi ^i\sigma ^{jk}\delta h_{jk}
\end{align}
and we know that a non-zero boundary charge only exists if $B^i$ is non-zero for the considered field configuration and transformation. However, since $\sigma ^{ij}=0=m_{ij}$ for the Kerroll black hole, the diffeomorphism charge vanishes for any $\xi ^i$. 

We may instead look at a similar quantity for the generator of subleading diffeomorphisms $D^{(1)}[\xi ]$ to find 
\begin{align}
    \delta D^{(1)}[\xi ]&= \int \dd ^dx\, \xi ^i \delta \mH _i^{(1)}=\int \dd ^dx\,\big[(...)\delta h_{ij}+(...)\delta \rho ^{ij}+ \partial _iB^i \big]
\end{align}
with 
\begin{align}
    B^i=-2\xi^{(j}\rho ^{k)i}\delta h_{jk}-2\xi_{j}\delta \rho ^{ji}+\xi ^i\rho ^{jk}\delta h_{jk}\,.
\end{align}
In this case, we obtain a non-vanishing boundary contribution and to render the generator functionally differentiable we have to cancel it by adding a boundary term $\delta Q^{(1)}_\xi $ to $\delta D^{(1)}[\xi ]$. On-shell this boundary term is just given by 
\begin{align}
    \delta Q^{(1)}_\xi =\lim _{r\to \infty} \oint \dd S_i \,\big(2\xi ^{(j} \rho ^{k)i}\delta h_{jk}+2\xi _j\delta \rho ^{ji}-\xi ^i\rho ^{jk}\delta h_{jk}\big) \quad \Rightarrow \quad \delta Q^{(1)}_{-\partial _\varphi }= \frac{\delta(r_\mathrm{s} a)}{2G_\mathrm{E}}
\end{align}
We therefore find a non-zero charge, $Q_{-\partial _\varphi }=Ma=:J^{(1)}$ associated to $-\partial _\varphi $ which sits in the subleading diffeomorphisms. We interpret this as another way of seeing that taking the strict Carroll limit of the Kerr metric cannot reproduce angular momentum. Instead, angular momentum is a subleading effect: Starting from the relativistic angular momentum $J$ in the Kerr metric \eqref{eq:kerr_mag_metric} one equivalently obtains $J^{(1)}$ by extracting the $\mathcal{O}(c)$-coefficient, $J=aE/c=J^{(1)}c$.

Finally, the boundary charge associated to the Killing vector of time translations $\partial _t$ may be computed by considering the generator associated to the scalar constraint
\begin{align}
    \delta G[\xi =\partial _t]=\int \dd ^dx\, N_{(0)}\delta \mH ^{(2)}=\int \dd ^dx\,\big[(...)\delta h_{ij}+(...)\delta \rho ^{ij}+ \partial _iB^i \big]
\end{align}
where we used that the canonical transformation parameters related with $\mH ^{(2)}$ and $\mH _i^{(2)}$ associated to a given diffeomorphism generated by $\xi ^\mu $ are given by $N_{(0)}\xi ^t$ and $\xi ^i+N_{(0)}^i\xi ^t $, respectively. The boundary term then reads 
\begin{align}
    B^i=-\frac{\sqrt{h}}{16\pi G_\mathrm{E}}\Big(2N_{(0)}h^{j[i}\nabla ^{k]} \delta h_{jk}-2\nabla ^{[i}N_{(0)}h^{j]k}\delta h_{jk}\Big) 
\end{align}
which implies the boundary charge 
\begin{align}
    \delta Q_{\partial _t} =\lim _{r\to \infty }\frac{1}{16\pi G_\mathrm{E}}\oint \dd S_i \sqrt{h}\Big(2N_{(0)}h^{j[i}\nabla ^{k]} \delta h_{jk}-2\nabla ^{[i}N_{(0)}h^{j]k}\delta h_{jk}\Big)=\frac{\delta r_\mathrm{s}}{2G_\mathrm{E}} 
\end{align}
confirming the expected relation between the mass parameter and energy, $Q_{\partial _t}=M$. Here we used that $r_{\mathrm{s}}=2G_\mathrm{E}M$. From the perspective of the relativistic parent theory this charge appears at $\mathcal{O}(c^2)$, i.e. the relativistic energy expands as $E=E^{(2)}c^2+\mathcal{O}(c^3)$. With $E^{(2)}=M$ this is consistent with keeping the mass parameter fixed.


\section{Conclusion and Outlook}\label{Conclusion}

In this article, we have presented two complementary constructions of rotating black holes in Carrollian gravity. The first resides in magnetic Carroll gravity, where we have shown that the Carrollian connection can be used to dress Carroll--Schwarzschild black holes with a rotational charge. This construction exploits genuinely Carrollian features of the theory and has no direct analog in the relativistic setting. The second construction is based on an odd-power expansion in the speed of light of general relativity, which allows us to identify a Carrollian analog of the Kerr black hole, the Kerroll black hole. In this case, rotation arises from subleading contributions in odd powers of the speed of light.

We finish with a few open points that, we believe, are interesting for further investigations.

Given the definition of Carroll black holes \cite{Ecker:2023uwm} there is a vital aspect that we did not address so far: The thermodynamic properties of these solutions. Indeed, in both cases it is not obvious whether a first law may be derived from first principles. For the rotating Carroll--Schwarzschild solution derived in Section \ref{sec:CSS} we found boundary charges for energy and angular momentum but for establishing a first law relation an independent derivation of the temperature of these geometries is needed. In analogy to the two-dimensional case one may attempt this by demanding the absence of conical singularities in a Wick rotated geometry, i.e., impose $t\sim t+\beta$ and compute the four-dimensional Carroll analog of the Euler character (see \cite{Figueroa-OFarrill:2022mcy} for a definition of quadratic curvature invariants). Analogously to the Euclidean case \cite{Liberati:1997sp,Ma:2003uj} demanding $\chi = \chi(D^2)\times \chi(S^2) = 2$ should yield a relation between $\beta $ and the solution parameters $r_\mathrm{s}$ and $J$. 

As far as the thermodynamics of the Kerroll black hole of Subsection \ref{sec:Kerroll_BHs} is concerned, the first law has to follow from expanding the Lorentzian counterpart: Since we are extracting the theory at $\mathcal{O}(c^2)$ in an expansion around the electric limit we also expect the thermodynamic quantities to obey a relation consistent with that order. In other words, at $\mathcal{O}(c^2)$ we should have 
\begin{align}\label{eq:firstlaw_guess}
    \delta E^{(2)}=(T\delta S)^{(2)}+\Omega ^{(1)}\delta J^{(1)}
\end{align}
where $E^{(2)}=M$ and $J^{(1)}=aM$ are the boundary charges associated to time translations and rotations (see Subsection \ref{subsec:bdy-charges}). The expansions of $S$, $T$ and $\Omega $ are not fixed a priory but we may at least check for consistency with the expansions of the Lorentzian quantities associated to the Kerr family. This yields 
\begin{align}
    S=c^2S^{(2)}=c^2\Big(\frac{A_\mathrm{c}}{4G_\mathrm{E}\hbar c}\Big), && T=T^{(0)}=\frac{\hbar c}{2\pi }\frac{r_\mathrm{c}-r_-}{2(r_\mathrm{c}^2+a^2)}, \\ \Omega =c\Omega ^{(1)}=c\frac{a}{r_\mathrm{c}^2+a^2}, && J=cJ^{(1)}=caM
\end{align}
where $A_\mathrm{c}=4\pi (r_\mathrm{c}^2+a^2)$ is the area of the Carroll extremal surface at $r=r_\mathrm{c}$ and $r_-=r_\mathrm{s}/2 -\sqrt{(r_\mathrm{s}/2)^2-a^2}$. Additionally, we assumed here that the combination $\hbar c$ is held constant, similarly to the tantum gravity limit \cite{Ecker:2024czh}. While these expansions are indeed consistent with \eqref{eq:firstlaw_guess} it would be interesting to see whether the expressions for $S^{(2)}$, $T^{(0)}$ and $\Omega ^{(1)}$ admit an independent derivation within extended magnetic Carroll gravity.

Interestingly, there is a tantum gravity-like \cite{Ecker:2024czh} multi scaling limit of Kerr black hole thermodynamics that leads to finite energy $E=Mc^2$, temperature $T$, entropy $S$, horizon radii $r_\pm$, Kerr parameter $a$, and (rescaled) horizon velocity $\Omega$/angular momentum $J$,
\begin{equation}
G_\mathrm{N}\to G_\mathrm{M}\,c^4\qquad\qquad \hbar\to\frac{\hbar}{c}\qquad\qquad \Omega\to\frac{\Omega}{c}\qquad\qquad J\to J\,c \qquad \qquad c\to 0\,.
\label{eq:tantengravitation}
\end{equation}
On the level of the metric this limit corresponds to the second option discussed in the Introduction after \eqref{eq:kerr_mag_metric}. It could be rewarding to find a Carroll theory where the scaling above appears naturally. Given that the scaling \eqref{eq:tantengravitation} keeps fixed the magnetic Carroll coupling $G_\mathrm{M}=G_\mathrm{N}/c^4$ rather than the electric Carroll coupling $G_\mathrm{E}=G_\mathrm{N}/c^2$, the extended magnetic Carroll gravity theory of Section \ref{sec:4} is not a candidate for such a theory. Neither is magnetic Carroll gravity because of the constraint violation $R[g_{ij}]\neq 0$.

In general, it would also be worthwhile to study Carroll extremal surfaces \cite{Ecker:2023uwm} in higher spacetime dimensions $D>2$ in more detail, particularly in the context of rotating black holes and extended magnetic Carroll gravity. Additionally, it would be gratifying to investigate whether there exists an extension or modification of general relativity for which extended magnetic Carroll gravity arises as a direct $c\to 0$ limit.

Finally, we have only considered the rotating Carroll--Schwarzschild black hole and the Carroll analog of the Kerr black hole in this article. It could be rewarding to study the possibly rich class of rotating solutions in magnetic Carroll gravity and to construct other Carroll analogs of Lorentzian black holes such as higher dimensional black holes, for instance Myers--Perry black holes \cite{Myers:1986un} in extended magnetic Carroll gravity.


\acknowledgments
\addcontentsline{toc}{section}{Acknowledgments}

We thank Ankit Aggarwal, Stefan Fredenhagen, Iva Lovrekovic, and Stefan Prohazka for discussions at ESI and/or the weekly Carroll group meetings at TU Wien.

FE, DG, LH, and AP thank the Erwin--Schr\"odinger Institute (ESI) for the hospitality in April 2024 during the program ``Carrollian physics and holography'' and we are grateful to the participants of this program for numerous insightful discussions on Carroll gravity and Carroll black holes.

FE and DG were supported by the Austrian Science Fund (FWF) [Grants DOI: \href{https://www.fwf.ac.at/en/research-radar/10.55776/P36619}{10.55776/P36619}, \href{https://www.fwf.ac.at/en/research-radar/10.55776/PAT1871425}{10.55776/PAT1871425}].

LH thanks the ETH Foundation for support through the Excellence Scholarship and Opportunity Programme.

ML-D thanks the Philippe Meyer Institute for financial support.

The research of AP is partially supported by the ANID Fondecyt grant 1260427. AP acknowledges
support from the Erwin Schrödinger Institute (ESI), where part of this research was carried out during the ESI Research in Teams Programme ``Exploring the infrared triangle beyond Lorentz symmetry''.


\appendix
\section{Poisson brackets of NNLO gravity}\label{sec:Poisson-brackets}
Here we collect the Poisson brackets of the theory obtained in Section \ref{sec:4} by expanding general relativity up to $\mathcal{O}(c^2)$ in odd powers of $c$. The non-vanishing scalar Hamiltonian Poisson brackets are
\begin{align}
  \{\mH^{(2)}(x),\mH^{(2)}(y)\} &=
  \mH_{(0)}^i(x)\,\partial_i\delta(x-y)
  + \mH_{(0)}^i(y)\,\partial_i\delta(x-y)
\end{align}
where \(\mH_{(0)}^i\equiv \glo^{ij}\mH^{(0)}_j\). The non-vanishing momentum Poisson brackets are
\begin{align}
  \{\mH^{(1)}_i(x),\mH^{(1)}_j(y)\} &=
    \mH^{(0)}_i(y)\,\partial_j\delta(x-y)
  + \mH^{(0)}_j(x)\,\partial_i\delta(x-y)\\
  \{\mH^{(2)}_i(x),\mH^{(0)}_j(y)\} &=
    \mH^{(0)}_i(y)\,\partial_j\delta(x-y)
  + \mH^{(0)}_j(x)\,\partial_i\delta(x-y)\\
  \{\mH^{(2)}_i(x),\mH^{(1)}_j(y)\} &=
    \mH^{(1)}_i(y)\,\partial_j\delta(x-y)
  + \mH^{(1)}_j(x)\,\partial_i\delta(x-y)\\
  \{\mH^{(2)}_i(x),\mH^{(2)}_j(y)\} &=
    \mH^{(2)}_i(y)\,\partial_j\delta(x-y)
  + \mH^{(2)}_j(x)\,\partial_i\delta(x-y)\,.
\end{align}
Finally, the non-vanishing mixed Hamiltonian-momentum brackets evaluate to
\begin{align}
  \{\mH^{(1)}_i(x),\mH^{(1)}(y)\} &= \mH^{(0)}(x)\,\partial_i\delta(x-y)\\
  \{\mH^{(1)}_i(x),\mH^{(2)}(y)\} &= \mH^{(1)}(x)\,\partial_i\delta(x-y)\\
  \{\mH^{(2)}_i(x),\mH^{(0)}(y)\} &= \mH^{(0)}(x)\,\partial_i\delta(x-y)\\
  \{\mH^{(2)}_i(x),\mH^{(1)}(y)\} &= \mH^{(1)}(x)\,\partial_i\delta(x-y)\\
  \{\mH^{(2)}_i(x),\mH^{(2)}(y)\} &= \mH^{(2)}(x)\,\partial_i\delta(x-y)\,.
\end{align}

\section{Canonical generators}\label{sec:Diff-generators}
It is useful to compute the momentum Poisson brackets above in a smeared-out form. For this we need the corresponding diffeomorphism generators of the theory. Start from the GR diffeomorphism generator which reads up to boundary terms
\begin{align}
    D[\xi] = \int \dd ^d x\, \xi ^i \mH _i = \int \dd^dx \,\pi^{ij}\mL_\xi g_{ij} ~.
\end{align}
Using the expansion from Section \ref{sec:4} we define
\begin{subequations}
    \begin{align}
    D^{(0)}[\xi] &= \int \dd ^d x\, \xi ^i \mH ^{(0)}_i =\int \dd^dx\,\pilo^{ij}\mL_\xi \glo_{ij} \\
    D^{(1)}[\xi]  &= \int \dd ^d x\, \xi ^i \mH ^{(1)}_i= \int \dd^dx\left(\pilo^{ij}\mL_\xi m_{ij}+\rho^{ij}\mL_\xi \glo_{ij}\right) \\
    D^{(2)}[\xi]  &= \int \dd ^d x\, \xi ^i \mH ^{(2)}_i= \int \dd^dx\left(\pilo^{ij}\mL_\xi l_{ij}+\rho^{ij}\mL_\xi m_{ij}+\sigma^{ij}\mL_\xi \glo_{ij}\right)\,. 
\end{align}
\end{subequations}
The action of the latter on the canonical fields in the full NNLO theory is
\begin{subequations}
    \begin{align}
    \{\glo_{ij},\,D^{(0)}[\xi]\} &= 0 & \{m_{ij},\,D^{(0)}[\xi]\}&=0 & \{l_{ij},\,D^{(0)}[\xi]\} &=\mL_\xi \glo_{ij} \\
    \{\pilo^{ij},\,D^{(0)}[\xi]\} &= 0 & \{\rho^{ij},\,D^{(0)}[\xi]\}&=0 &  \{\sigma^{ij},\,D^{(0)}[\xi]\} &=\mL_\xi \pilo^{ij}\\
    \{\glo_{ij},\,D^{(1)}[\xi]\} &=0 & \{m_{ij},\,D^{(1)}[\xi]\}&=\mL_\xi \glo_{ij} & \{l_{ij},\,D^{(1)}[\xi]\}&=\mL_\xi m_{ij} \\
    \{\pilo^{ij},\,D^{(1)}[\xi]\} &= 0 & \{\rho^{ij},\,D^{(1)}[\xi]\}&=\mL_\xi\pilo^{ij} &  \{\sigma^{ij},\,D^{(1)}[\xi]\}&=\mL_\xi \rho^{ij} \\
    \{\glo_{ij},\,D^{(2)}[\xi]\}&=\mL_\xi \glo_{ij} &  \{m_{ij},\,D^{(2)}[\xi]\}&=\mL_\xi m_{ij} & \{l_{ij},\,D^{(2)}[\xi]\}&=\mL_\xi l_{ij} \\
    \{\pilo^{ij},\,D^{(2)}[\xi]\} &= \mL_\xi\pilo^{ij} & \{\rho^{ij},\,D^{(2)}[\xi]\}&=\mL_\xi\rho^{ij} &  \{\sigma^{ij},\,D^{(2)}[\xi]\}&=\mL_\xi \sigma^{ij}\,.
\end{align}
\end{subequations}
We can directly obtain the Poisson brackets of extended magnetic Carroll gravity by setting $\pilo^{ij}=0$, which essentially means that we can omit all appearances of $l_{ij}$ and $\pilo^{ij}$ in the Poisson brackets. The diffeomorphism generators change accordingly.

\section{Odd-power Carroll gravity expansion in second-order formalism}\label{sec:appendix-2nd-order}
Here, we briefly sketch how to perform an odd-power expansion of general relativity along the lines of \cite{Hansen:2021fxi}. In order to generalize the Carroll expansion of Einstein gravity to include odd powers in $c$, we start from the established PUL decomposition
\begin{align}\label{eq:PUL-split}
    g_{\mu\nu}&=\Pi_{\mu\nu}-c^2T_\mu T_\nu & g^{\mu\nu}&=\Pi^{\mu\nu}-\frac{1}{c^2}V^\mu V^\nu
\end{align}
with
\begin{align}\label{eq:PUL-geometry}
    V^\mu T_\mu &= -1 & \Pi_{\mu\nu}V^\nu &=0 & \Pi^{\mu\nu} T_\nu &= 0 & \delta^\mu_\nu &= -V^\mu T_\nu + \Pi^{\mu\rho}\Pi_{\rho\nu}\,.
\end{align}
In the context of PUL variables $\Pi^{\mu\nu}$ denotes the (inverse) spatial metric, not the leading order canonical momentum, which has spatial indices $\Pi^{ij}$.
We then consider a more general expansion than in \cite{Hansen:2021fxi}, including subleading contributions in odd powers of $c$
\begin{align}
    T_\mu &= \tau_\mu + c t_\mu + c^2 s_\mu + \mO(c^3) &
    \Pi_{\mu \nu} &= \glo_{\mu \nu} + c m_{\mu \nu} + c^2 l_{\mu \nu} + \mO(c^3) \\
    V^\mu &= v^\mu + c u^\mu + c^2 w^\mu + \mO(c^3) &
    \Pi^{\mu \nu} &= h^{\mu \nu} + c m^{\mu \nu} + c^2 l^{\mu \nu} + \mO(c^3)\,.
\end{align}
By setting the odd-power fields to zero, we recover the even-power expansion considered in \cite{Hansen:2021fxi}. The kinematic relations between these fields are fixed by expanding \eqref{eq:PUL-geometry}
order by order, which yields constraints for the subleading fields, similarly to those for Carroll geometry
\begin{align}
    \tau_\mu v^\mu &= -1 & h_{\mu\nu}v^\mu &=0 & h^{\mu\nu} \tau_\nu &= 0 & \delta^\mu_\nu &= -v^\mu \tau_\nu + h^{\mu\rho} h_{\rho\nu}\,.
\end{align} 
Including subleading odd-power contribution necessitates an expansion of the extrinsic curvature beyond the leading order quantity $K_{\mu\nu} = -\frac{1}{2}\mL_v h_{\mu\nu}$ as follows
\begin{multline}
    \mK_{\mu\nu} =-\frac{1}{2}\mL_V \Pi_{\mu\nu} =-\frac{1}{2}\mL_{v+c u+c^2w+\mO(c^3)} \left( h_{\mu\nu} + c m_{\mu\nu} + c^2  l_{\mu\nu}+\mO(c^3)\right) \\ = K_{\mu\nu} + c S_{\mu\nu} + c^2 L_{\mu\nu}+\mO(c^3)
\end{multline}
From $V^\mu \mK_{\mu\nu} = 0$ we obtain
\begin{align} \label{eq:extrinsic-curvature-kinematics}
    v^\mu K_{\mu\nu} &= 0 & v^\mu S_{\mu\nu} + u^\mu K_{\mu\nu} &= 0 & w^\mu K_{\mu\nu}+u^\mu S_{\mu\nu} + v^\mu L_{\mu\nu} &= 0\,.
\end{align}
In general $v^\mu S_{\mu\nu} \neq 0$ and $v^\mu L_{\mu\nu}\neq 0$, which means that we cannot use $h^{\mu\nu}$ as a metric-like object to move the indices of $S_{\mu\nu}$ and $L_{\mu\nu}$. Instead, it is more convenient to separately expand
\begin{align}
    \mathcal{K}^{\mu\nu} =\mK_{\rho\sigma} \Pi^{\rho\mu}\Pi^{\sigma\nu}= K^{\mu\nu} + c S^{\mu\nu} + c^2 L^{\mu\nu}\,.
\end{align}
Moreover, we write $\mK = \mK_{\mu\nu} \Pi^{\mu\nu}$ as
\begin{align}
    \mK = K+c S+ c^2L\,.
\end{align}
Let us now turn to the odd-power expansion of the Einstein--Hilbert action in PUL variables
\begin{align} \label{eq:EH-action-in-PUL-variables}
 I_\mathrm{EH}=\frac{c^2}{16\pi G}
  \int_M\!\dd^{d+1}x E\,\big[
    \big(\mathcal{K}^{\mu\nu} \mathcal{K}_{\mu\nu} - \mathcal{K}^2\big)
    + c^2 \Pi^{\mu\nu} \mathcal{R}_{\mu\nu}
    + \frac{c^4}{4} \Pi^{\mu\nu} \Pi^{\rho\sigma}
    \left(dT\right)_{\mu\rho} \left(dT\right)_{\nu\sigma}
  \big]
\end{align}
where $E=\sqrt{\det\left(T_\mu T_\nu + \Pi_{\mu\nu}\right)}$ denotes the PUL vielbein determinant. We note that in a systematic expansion in $c$, the equations of motion of the lower order action is always encoded in the higher order action \cite{Hansen:2020pqs}. Writing
\begin{align}
    I_\mathrm{EH} = I_0 + c I_1 + c^2 I_2 + \mO(c^3)
\end{align}
we find
\begin{subequations}
    \begin{align}\label{eq:electric-gravity-action-as-from-odd-power-expansion}
    I_\mathrm{LO}= \frac{1}{16\pi G_\mathrm{E}} \int \dd^{d+1}x\mathfrak{e}\,\big( K^{\mu\nu}K_{\mu\nu} - K^2\big)
\end{align}
\begin{align}\label{eq:NL-action-from-odd-power-exp}
    I_\mathrm{NLO} = \frac{1}{16\pi G_\mathrm{E}} \int \dd^{d+1}x\mathfrak{e}\, \Big( S_{\mu\nu}K^{\mu\nu}+S^{\mu\nu}K_{\mu\nu}  -2 KS+\frac{\mathfrak{e}^{(1)}}{\mathfrak{e}}\big(K^{\mu\nu}K_{\mu\nu} - K^2\big)\Big)
\end{align}
\begin{multline}\label{eq:NNL-action-from-odd-power-exp}
    I_\mathrm{NNLO}= \frac{1}{16\pi G_\mathrm{E}} \int \dd^{d+1}x\mathfrak{e}\, \Big(h^{\mu\nu}R_{\mu\nu}+ S^{\mu\nu}S_{\mu\nu}+ L_{\mu\nu}K^{\mu\nu}+L^{\mu\nu}K_{\mu\nu}  -S^2-2 KL  \\ +\frac{\mathfrak{e}^{(1)}}{\mathfrak{e}}\left(S^{\mu\nu}K_{\mu\nu}+S_{\mu\nu}K^{\mu\nu}  -2 KS\right)+\frac{\mathfrak{e}^{(2)}}{\mathfrak{e}}\big(K^{\mu\nu}K_{\mu\nu} - K^2\big)\Big)
\end{multline}
\end{subequations}
where $\mathfrak{e}^{(1)}$ and $\mathfrak{e}^{(2)}$ stem from the expansion of the vielbein determinant $E=\mathfrak{e}+c \mathfrak{e}^{(1)}+c^2 \mathfrak{e}^{(2)}+\mO(c^3)$. We observe that the leading order action again reproduces electric Carroll gravity, whereas the odd-power NLO term is new and by construction does not feature in the even power expansion at all. The NNLO term picks up additional contributions from the odd power fields compared to \cite{Hansen:2021fxi}, which are crucial to realize the rotating Kerroll black hole solution.


\renewcommand{\emph}{\textit}
\renewcommand{\em}{\it}

\addcontentsline{toc}{section}{References}
\bibliographystyle{utphys} 
\bibliography{review} 

@article{Bal:2026xup,
    author = "Bal, Enes and Ekiz, Ertu{\u{g}}rul and Kahya, Emre Onur and Zorba, Utku",
    title = "{Stationary solutions in the small-$c$ expansion of GR}",
    eprint = "2604.23677",
    archivePrefix = "arXiv",
    primaryClass = "gr-qc",
    month = "4",
    year = "2026"
}

@article{Bagchi:2023cen,
    author = "Bagchi, Arjun and Dhivakar, Prateksh and Dutta, Sudipta",
    title = "{Holography in flat spacetimes: the case for Carroll}",
    eprint = "2311.11246",
    archivePrefix = "arXiv",
    primaryClass = "hep-th",
    doi = "10.1007/JHEP08(2024)144",
    journal = "JHEP",
    volume = "08",
    pages = "144",
    year = "2024"
}

@article{Liberati:1997sp,
    author = "Liberati, Stefano and Pollifrone, Giuseppe",
    title = "{Entropy and topology for gravitational instantons}",
    eprint = "hep-th/9708014",
    archivePrefix = "arXiv",
    reportNumber = "CERN-TH-97-116",
    doi = "10.1103/PhysRevD.56.6458",
    journal = "Phys. Rev. D",
    volume = "56",
    pages = "6458--6466",
    year = "1997"
}

@article{Ma:2003uj,
    author = "Ma, Zheng Ze",
    title = "{Euler numbers of four-dimensional rotating black holes with the Euclidean signature}",
    eprint = "hep-th/0512127",
    archivePrefix = "arXiv",
    doi = "10.1103/PHYSREVD.67.024027",
    journal = "Phys. Rev. D",
    volume = "67",
    pages = "024027",
    year = "2003"
}

@article{Myers:1986un,
    author = "Myers, Robert C. and Perry, M. J.",
    title = "{Black Holes in Higher Dimensional Space-Times}",
    reportNumber = "PRINT-86-0067 (PRINCETON)",
    doi = "10.1016/0003-4916(86)90186-7",
    journal = "Annals Phys.",
    volume = "172",
    pages = "304",
    year = "1986"
}

@article{Hansen:2020wqw,
    author = "Hansen, Dennis and Hartong, Jelle and Obers, Niels A. and Oling, Gerben",
    title = "{Galilean first-order formulation for the nonrelativistic expansion of general relativity}",
    eprint = "2012.01518",
    archivePrefix = "arXiv",
    primaryClass = "hep-th",
    reportNumber = "NORDITA 2020-112",
    doi = "10.1103/PhysRevD.104.L061501",
    journal = "Phys. Rev. D",
    volume = "104",
    number = "6",
    pages = "L061501",
    year = "2021"
}

@article{Musaeus:2023oyp,
    author = "Musaeus, J{\o}rgen and Obers, Niels A. and Oling, Gerben",
    title = "{Setting the connection free in the Galilei and Carroll expansions of gravity}",
    eprint = "2312.13924",
    archivePrefix = "arXiv",
    primaryClass = "hep-th",
    reportNumber = "NORDITA 2023-093",
    doi = "10.1103/PhysRevD.109.104040",
    journal = "Phys. Rev. D",
    volume = "109",
    number = "10",
    pages = "104040",
    year = "2024"
}

@article{Bagchi:2024rje,
    author = "Bagchi, Arjun and Banerjee, Aritra and Hartong, Jelle and Have, Emil and Kolekar, Kedar S.",
    title = "{Strings near black holes are Carrollian. Part II}",
    eprint = "2407.12911",
    archivePrefix = "arXiv",
    primaryClass = "hep-th",
    doi = "10.1007/JHEP11(2024)024",
    journal = "JHEP",
    volume = "11",
    pages = "024",
    year = "2024"
}

@article{Ahmadi-Jahmani:2025iqc,
    author = "Ahmadi-Jahmani, M. M. and Parvizi, A.",
    title = "{Fracton and non-Lorentzian particle duality: gauge field couplings and geometric implications}",
    eprint = "2503.21660",
    archivePrefix = "arXiv",
    primaryClass = "hep-th",
    doi = "10.1007/JHEP08(2025)157",
    journal = "JHEP",
    volume = "08",
    pages = "157",
    year = "2025"
}

@phdthesis{Ecker:2025vnl,
    author = "Ecker, Florian",
    title = "{Carroll symmetries in field theory and gravity}",
    eprint = "2603.12902",
    archivePrefix = "arXiv",
    primaryClass = "hep-th",
    school = "TU. Vienna",
    month = "3",
    year = "2026"
}

@article{Ruzziconi:2026bix,
    author = "Ruzziconi, Romain",
    title = "{Carrollian physics and holography}",
    eprint = "2602.02644",
    archivePrefix = "arXiv",
    primaryClass = "hep-th",
    doi = "10.1016/j.physrep.2026.03.005",
    journal = "Phys. Rept.",
    volume = "1182",
    pages = "1--87",
    year = "2026"
}

@article{Kolar:2025ebv,
    author = "Kol{\'a}{\v{r}}, Ivan and Kubiznak, David and Tadros, Poula",
    title = "{Rotating Carroll black holes: A no go theorem}",
    eprint = "2506.10451",
    archivePrefix = "arXiv",
    primaryClass = "hep-th",
    reportNumber = "APS/123-QED",
    doi = "10.1103/twv1-kphf",
    journal = "Phys. Rev. D",
    volume = "112",
    number = "12",
    pages = "L121504",
    year = "2025"
}

@article{Pekar:2024ukc,
    author = "Pekar, Simon and P{\'e}rez, Alfredo and Salgado-Rebolledo, Patricio",
    title = "{Cartan-like formulation of electric Carrollian gravity}",
    eprint = "2406.01665",
    archivePrefix = "arXiv",
    primaryClass = "hep-th",
    doi = "10.1007/JHEP09(2024)059",
    journal = "JHEP",
    volume = "09",
    pages = "059",
    year = "2024"
}

@article{Tadros:2024bev,
    author = "Tadros, Poula and Kol{\'a}{\v{r}}, Ivan",
    title = "{Carroll black holes in (A)dS spacetimes and their higher-derivative modifications}",
    eprint = "2408.01836",
    archivePrefix = "arXiv",
    primaryClass = "gr-qc",
    doi = "10.1103/PhysRevD.110.084064",
    journal = "Phys. Rev. D",
    volume = "110",
    number = "8",
    pages = "084064",
    year = "2024"
}

@article{Fiorucci:2025twa,
    author = "Fiorucci, Adrien and Pekar, Simon and Marios Petropoulos, P. and Vilatte, Matthieu",
    title = "{Carrollian-Holographic Derivation of Gravitational Flux-Balance Laws}",
    eprint = "2505.00077",
    archivePrefix = "arXiv",
    primaryClass = "hep-th",
    reportNumber = "CPHT-RR035.042025",
    doi = "10.1103/qv17-ks32",
    journal = "Phys. Rev. Lett.",
    volume = "135",
    number = "26",
    pages = "261602",
    year = "2025"
}

@article{Bagchi:2016yyf,
    author = "Bagchi, Arjun and Chakrabortty, Shankhadeep and Parekh, Pulastya",
    title = "{Tensionless Superstrings: View from the Worldsheet}",
    eprint = "1606.09628",
    archivePrefix = "arXiv",
    primaryClass = "hep-th",
    reportNumber = "MIT-CTP-4816",
    doi = "10.1007/JHEP10(2016)113",
    journal = "JHEP",
    volume = "10",
    pages = "113",
    year = "2016"
}

@article{Bagchi:2025vri,
    author = "Bagchi, Arjun and Banerjee, Aritra and Dhivakar, Prateksh and Mondal, Saikat and Shukla, Ashish",
    title = "{The Carrollian Kaleidoscope}",
    eprint = "2506.16164",
    archivePrefix = "arXiv",
    primaryClass = "hep-th",
    month = "6",
    year = "2025"
}

@article{Poulias:2025eck,
    author = "Poulias, Georgios and Vandoren, Stefan",
    title = "{On Carroll partition functions and flat space holography}",
    eprint = "2503.20615",
    archivePrefix = "arXiv",
    primaryClass = "hep-th",
    doi = "10.1007/JHEP06(2025)232",
    journal = "JHEP",
    volume = "06",
    pages = "232",
    year = "2025"
}

@article{Aviles:2025ygw,
    author = "Avil{\'e}s, Luis and Fuentealba, Oscar and Hidalgo, Diego and Rodr{\'\i}guez, Pablo",
    title = "{AdS$_{3}$ Carroll gravity: asymptotic symmetries and C-thermal configurations}",
    eprint = "2503.18818",
    archivePrefix = "arXiv",
    primaryClass = "hep-th",
    doi = "10.1007/JHEP05(2025)174",
    journal = "JHEP",
    volume = "05",
    pages = "174",
    year = "2025"
}

@article{Ecker:2024czh,
    author = "Ecker, Florian and Fiorucci, Adrien and Grumiller, Daniel",
    title = "{Tantum gravity}",
    eprint = "2501.00095",
    archivePrefix = "arXiv",
    primaryClass = "hep-th",
    doi = "10.1103/PhysRevD.111.L021901",
    journal = "Phys. Rev. D",
    volume = "111",
    number = "2",
    pages = "L021901",
    year = "2025"
}

@article{Hansen:2020pqs,
    author = "Hansen, Dennis and Hartong, Jelle and Obers, Niels A.",
    title = "{Non-Relativistic Gravity and its Coupling to Matter}",
    eprint = "2001.10277",
    archivePrefix = "arXiv",
    primaryClass = "gr-qc",
    doi = "10.1007/JHEP06(2020)145",
    journal = "JHEP",
    volume = "06",
    pages = "145",
    year = "2020"
}

@article{Bagchi:2024qsb,
    author = "Bagchi, Arjun and Chakraborty, Pronoy and Chakrabortty, Shankhadeep and Fredenhagen, Stefan and Grumiller, Daniel and Pandit, Priyadarshini",
    title = "{Boundary Carrollian Conformal Field Theories and Open Null Strings}",
    eprint = "2409.01094",
    archivePrefix = "arXiv",
    primaryClass = "hep-th",
    reportNumber = "TUW-24-05",
    doi = "10.1103/PhysRevLett.134.071604",
    journal = "Phys. Rev. Lett.",
    volume = "134",
    number = "7",
    pages = "071604",
    year = "2025"
}

@article{Grumiller:2024dql,
    author = "Grumiller, Daniel and Montecchio, Luciano and Nejati, Mohaddese Shams",
    title = "{Carroll dilaton supergravity in two dimensions}",
    eprint = "2409.17781",
    archivePrefix = "arXiv",
    primaryClass = "hep-th",
    reportNumber = "TUW-24-03",
    doi = "10.1007/JHEP12(2024)005",
    journal = "JHEP",
    volume = "12",
    pages = "005",
    year = "2024"
}

@article{Bagchi:2023cfp,
    author = "Bagchi, Arjun and Banerjee, Aritra and Hartong, Jelle and Have, Emil and Kolekar, Kedar S. and Mandlik, Mangesh",
    title = "{Strings near black holes are Carrollian}",
    eprint = "2312.14240",
    archivePrefix = "arXiv",
    primaryClass = "hep-th",
    doi = "10.1103/PhysRevD.110.086009",
    journal = "Phys. Rev. D",
    volume = "110",
    number = "8",
    pages = "086009",
    year = "2024"
}

@article{Pena-Benitez:2023aat,
    author = "Pe{\~n}a-Ben{\'\i}tez, Francisco and Salgado-Rebolledo, Patricio",
    title = "{Fracton gauge fields from higher-dimensional gravity}",
    eprint = "2310.12610",
    archivePrefix = "arXiv",
    primaryClass = "hep-th",
    doi = "10.1007/JHEP04(2024)009",
    journal = "JHEP",
    volume = "04",
    pages = "009",
    year = "2024"
}

@article{Isham:1975ur,
    author = "Isham, C. J.",
    title = "{Some Quantum Field Theory Aspects of the Superspace Quantization of General Relativity}",
    reportNumber = "Print-76-0255 (KING S COLL.)",
    doi = "10.1098/rspa.1976.0138",
    journal = "Proc. Roy. Soc. Lond. A",
    volume = "351",
    pages = "209--232",
    year = "1976"
}

@inproceedings{Teitelboim:1978wv,
    author = "Teitelboim, Claudio",
    title = "{Surface deformations, their Square Root and the Signature of Spacetime}",
    booktitle = "{7th International Group Theory Colloquium: The Integrative Conference on Group Theory and Mathematical Physics}",
    reportNumber = "Print-78-1134 (PRINCETON)",
    month = "12",
    year = "1978"
}

@article{Matulich:2019cdo,
    author = "Matulich, Javier and Prohazka, Stefan and Salzer, Jakob",
    title = "{Limits of three-dimensional gravity and metric kinematical Lie algebras in any dimension}",
    eprint = "1903.09165",
    archivePrefix = "arXiv",
    primaryClass = "hep-th",
    doi = "10.1007/JHEP07(2019)118",
    journal = "JHEP",
    volume = "07",
    pages = "118",
    year = "2019"
}

@article{Bagchi:2022iqb,
    author = "Bagchi, Arjun and Grumiller, Daniel and Sheikh-Jabbari, Shahin and Sheikh-Jabbari, M. M.",
    title = "{Horizon strings as 3D black hole microstates}",
    eprint = "2210.10794",
    archivePrefix = "arXiv",
    primaryClass = "hep-th",
    reportNumber = "TUW-22-05",
    doi = "10.21468/SciPostPhys.15.5.210",
    journal = "SciPost Phys.",
    volume = "15",
    number = "5",
    pages = "210",
    year = "2023"
}

@article{Bergshoeff:2024ilz,
    author = "Bergshoeff, Eric A. and Concha, Patrick and Fierro, Octavio and Rodr{\'\i}guez, Evelyn and Rosseel, Jan",
    title = "{A conformal approach to Carroll gravity}",
    eprint = "2412.17752",
    archivePrefix = "arXiv",
    primaryClass = "hep-th",
    doi = "10.1007/JHEP07(2025)075",
    journal = "JHEP",
    volume = "07",
    pages = "075",
    year = "2025"
}

@article{Redondo-Yuste:2022czg,
    author = "Redondo-Yuste, Jaime and Lehner, Luis",
    title = "{Non-linear black hole dynamics and Carrollian fluids}",
    eprint = "2212.06175",
    archivePrefix = "arXiv",
    primaryClass = "gr-qc",
    doi = "10.1007/JHEP02(2023)240",
    journal = "JHEP",
    volume = "02",
    pages = "240",
    year = "2023"
}

@article{Bicak:2023rsz,
    author = "Bi\v{c}\'ak, Ji\v{r}\'\i{} and Kubiz\v{n}\'ak, David and Perche, T. Rick",
    title = "{Migrating Carrollian particles on magnetized black hole horizons}",
    eprint = "2302.11639",
    archivePrefix = "arXiv",
    primaryClass = "gr-qc",
    doi = "10.1103/PhysRevD.107.104014",
    journal = "Phys. Rev. D",
    volume = "107",
    number = "10",
    pages = "104014",
    year = "2023"
}

@article{Gray:2022svz,
    author = "Gray, Finnian and Kubiznak, David and Perche, T. Rick and Redondo-Yuste, Jaime",
    title = "{Carrollian motion in magnetized black hole horizons}",
    eprint = "2211.13695",
    archivePrefix = "arXiv",
    primaryClass = "gr-qc",
    doi = "10.1103/PhysRevD.107.064009",
    journal = "Phys. Rev. D",
    volume = "107",
    number = "6",
    pages = "064009",
    year = "2023"
}

@article{Bekaert:2015xua,
    author = "Bekaert, Xavier and Morand, Kevin",
    title = "{Connections and dynamical trajectories in generalised Newton-Cartan gravity II. An ambient perspective}",
    eprint = "1505.03739",
    archivePrefix = "arXiv",
    primaryClass = "hep-th",
    doi = "10.1063/1.5030328",
    journal = "J. Math. Phys.",
    volume = "59",
    number = "7",
    pages = "072503",
    year = "2018"
}

@article{Morand:2018tke,
    author = "Morand, Kevin",
    title = "{Embedding Galilean and Carrollian geometries I. Gravitational waves}",
    eprint = "1811.12681",
    archivePrefix = "arXiv",
    primaryClass = "hep-th",
    doi = "10.1063/1.5130907",
    journal = "J. Math. Phys.",
    volume = "61",
    number = "8",
    pages = "082502",
    year = "2020"
}

@article{Penna:2018gfx,
    author = "Penna, Robert F.",
    title = "{Near-horizon Carroll symmetry and black hole Love numbers}",
    eprint = "1812.05643",
    archivePrefix = "arXiv",
    primaryClass = "hep-th",
    month = "12",
    year = "2018"
}

@article{Gupta:2020dtl,
    author = "Gupta, Nishant and Suryanarayana, Nemani V.",
    title = "{Constructing Carrollian CFTs}",
    eprint = "2001.03056",
    archivePrefix = "arXiv",
    primaryClass = "hep-th",
    doi = "10.1007/JHEP03(2021)194",
    journal = "JHEP",
    volume = "03",
    pages = "194",
    year = "2021"
}

@article{Henneaux:1979vn,
    author = "Henneaux, Marc",
    title = "{Geometry of Zero Signature Space-times}",
    reportNumber = "PRINT-79-0606 (PRINCETON)",
    journal = "Bull. Soc. Math. Belg.",
    volume = "31",
    pages = "47--63",
    year = "1979"
}

@article{Ciambelli:2023tzb,
    author = "Ciambelli, Luca and Grumiller, Daniel",
    title = "{Carroll geodesics}",
    eprint = "2311.04112",
    archivePrefix = "arXiv",
    primaryClass = "hep-th",
    reportNumber = "TUW-23-05",
    doi = "10.1140/epjc/s10052-024-13232-4",
    journal = "Eur. Phys. J. C",
    volume = "84",
    number = "9",
    pages = "933",
    year = "2024"
}

@article{Grumiller:2017sjh,
    author = "Grumiller, Daniel and Merbis, Wout and Riegler, Max",
    title = "{Most general flat space boundary conditions in three-dimensional Einstein gravity}",
    eprint = "1704.07419",
    archivePrefix = "arXiv",
    primaryClass = "hep-th",
    reportNumber = "TUW-17-04",
    doi = "10.1088/1361-6382/aa8004",
    journal = "Class. Quant. Grav.",
    volume = "34",
    number = "18",
    pages = "184001",
    year = "2017"
}

@article{Ecker:2023uwm,
    author = "Ecker, Florian and Grumiller, Daniel and Hartong, Jelle and P\'erez, Alfredo and Prohazka, Stefan and Troncoso, Ricardo",
    title = "{Carroll black holes}",
    eprint = "2308.10947",
    archivePrefix = "arXiv",
    primaryClass = "hep-th",
    reportNumber = "TUW-23-03",
    doi = "10.21468/SciPostPhys.15.6.245",
    journal = "SciPost Phys.",
    volume = "15",
    number = "6",
    pages = "245",
    year = "2023"
}

@article{Figueroa-OFarrill:2023qty,
    author = "Figueroa-O'Farrill, Jos\'e and P\'erez, Alfredo and Prohazka, Stefan",
    title = "{Quantum Carroll/fracton particles}",
    eprint = "2307.05674",
    archivePrefix = "arXiv",
    primaryClass = "hep-th",
    reportNumber = "EMPG-23-13",
    doi = "10.1007/JHEP10(2023)041",
    journal = "JHEP",
    volume = "10",
    pages = "041",
    year = "2023"
}

@article{Figueroa-OFarrill:2023vbj,
    author = "Figueroa-O'Farrill, Jos\'e and P\'erez, Alfredo and Prohazka, Stefan",
    title = "{Carroll/fracton particles and their correspondence}",
    eprint = "2305.06730",
    archivePrefix = "arXiv",
    primaryClass = "hep-th",
    reportNumber = "EMPG-23-09",
    doi = "10.1007/JHEP06(2023)207",
    journal = "JHEP",
    volume = "06",
    pages = "207",
    year = "2023"
}

@article{Hansen:2021fxi,
    author = "Hansen, Dennis and Obers, Niels A. and Oling, Gerben and S\o{}gaard, Benjamin T.",
    title = "{Carroll Expansion of General Relativity}",
    eprint = "2112.12684",
    archivePrefix = "arXiv",
    primaryClass = "hep-th",
    reportNumber = "NORDITA 2021-156",
    doi = "10.21468/SciPostPhys.13.3.055",
    journal = "SciPost Phys.",
    volume = "13",
    number = "3",
    pages = "055",
    year = "2022"
}

@article{deBoer:2023fnj,
    author = "de Boer, Jan and Hartong, Jelle and Obers, Niels A. and Sybesma, Watse and Vandoren, Stefan",
    title = "{Carroll stories}",
    eprint = "2307.06827",
    archivePrefix = "arXiv",
    primaryClass = "hep-th",
    reportNumber = "NORDITA-2023-036",
    doi = "10.1007/JHEP09(2023)148",
    journal = "JHEP",
    volume = "09",
    pages = "148",
    year = "2023"
}

@article{Bergshoeff:2022eog,
    author = "Bergshoeff, Eric and Figueroa-O'Farrill, Jos\'e and Gomis, Joaquim",
    title = "{A non-lorentzian primer}",
    eprint = "2206.12177",
    archivePrefix = "arXiv",
    primaryClass = "hep-th",
    reportNumber = "EMPG-22-08",
    doi = "10.21468/SciPostPhysLectNotes.69",
    journal = "SciPost Phys. Lect. Notes",
    volume = "69",
    pages = "1",
    year = "2023"
}

@article{Bagchi:2022emh,
    author = "Bagchi, Arjun and Banerjee, Shamik and Basu, Rudranil and Dutta, Sudipta",
    title = "{Scattering Amplitudes: Celestial and Carrollian}",
    eprint = "2202.08438",
    archivePrefix = "arXiv",
    primaryClass = "hep-th",
    doi = "10.1103/PhysRevLett.128.241601",
    journal = "Phys. Rev. Lett.",
    volume = "128",
    number = "24",
    pages = "241601",
    year = "2022"
}

@article{Henneaux:2021yzg,
    author = "Henneaux, Marc and Salgado-Rebolledo, Patricio",
    title = "{Carroll contractions of Lorentz-invariant theories}",
    eprint = "2109.06708",
    archivePrefix = "arXiv",
    primaryClass = "hep-th",
    doi = "10.1007/JHEP11(2021)180",
    journal = "JHEP",
    volume = "11",
    pages = "180",
    year = "2021"
}

@article{Perez:2022jpr,
    author = "P\'erez, Alfredo",
    title = "{Asymptotic symmetries in Carrollian theories of gravity with a negative cosmological constant}",
    eprint = "2202.08768",
    archivePrefix = "arXiv",
    primaryClass = "hep-th",
    reportNumber = "CECS-PHY-21/04",
    doi = "10.1007/JHEP09(2022)044",
    journal = "JHEP",
    volume = "09",
    pages = "044",
    year = "2022"
}

@article{Guerrieri:2021cdz,
    author = "Guerrieri, Amanda and Sobreiro, Rodrigo F.",
    title = "{Carroll limit of four-dimensional gravity theories in the first order formalism}",
    eprint = "2107.10129",
    archivePrefix = "arXiv",
    primaryClass = "gr-qc",
    doi = "10.1088/1361-6382/ac345f",
    journal = "Class. Quant. Grav.",
    volume = "38",
    number = "24",
    pages = "245003",
    year = "2021"
}

@article{Figueroa-OFarrill:2022mcy,
    author = "Figueroa-O'Farrill, Jos\'e and Have, Emil and Prohazka, Stefan and Salzer, Jakob",
    title = "{The gauging procedure and carrollian gravity}",
    eprint = "2206.14178",
    archivePrefix = "arXiv",
    primaryClass = "hep-th",
    reportNumber = "EMPG-22-10",
    doi = "10.1007/JHEP09(2022)243",
    journal = "JHEP",
    volume = "09",
    pages = "243",
    year = "2022"
}

@article{Baig:2023yaz,
    author = "Baig, Saba Asif and Distler, Jacques and Karch, Andreas and Raz, Amir and Sun, Hao-Yu",
    title = "{Spacetime Subsystem Symmetries}",
    eprint = "2303.15590",
    archivePrefix = "arXiv",
    primaryClass = "hep-th",
    reportNumber = "UTWI-07-2023",
    month = "3",
    year = "2023"
}

@article{Marsot:2022imf,
    author = "Marsot, L. and Zhang, P. -M. and Chernodub, M. and Horvathy, P. A.",
    title = "{Hall effects in Carroll dynamics}",
    eprint = "2212.02360",
    archivePrefix = "arXiv",
    primaryClass = "hep-th",
    doi = "10.1016/j.physrep.2023.07.007",
    journal = "Phys. Rept.",
    volume = "1028",
    pages = "1--60",
    year = "2023"
}

@article{Bagchi:2016bcd,
    author = "Bagchi, Arjun and Basu, Rudranil and Kakkar, Ashish and Mehra, Aditya",
    title = "{Flat Holography: Aspects of the dual field theory}",
    eprint = "1609.06203",
    archivePrefix = "arXiv",
    primaryClass = "hep-th",
    doi = "10.1007/JHEP12(2016)147",
    journal = "JHEP",
    volume = "12",
    pages = "147",
    year = "2016"
}

@article{Donnay:2019jiz,
    author = "Donnay, Laura and Marteau, Charles",
    title = "{Carrollian Physics at the Black Hole Horizon}",
    eprint = "1903.09654",
    archivePrefix = "arXiv",
    primaryClass = "hep-th",
    doi = "10.1088/1361-6382/ab2fd5",
    journal = "Class. Quant. Grav.",
    volume = "36",
    number = "16",
    pages = "165002",
    year = "2019"
}

@article{Bergshoeff:2017btm,
    author = "Bergshoeff, Eric and Gomis, Joaquim and Rollier, Blaise and Rosseel, Jan and ter Veldhuis, Tonnis",
    title = "{Carroll versus Galilei Gravity}",
    eprint = "1701.06156",
    archivePrefix = "arXiv",
    primaryClass = "hep-th",
    doi = "10.1007/JHEP03(2017)165",
    journal = "JHEP",
    volume = "03",
    pages = "165",
    year = "2017"
}

@article{Levy1965,
  author =        {L{\'e}vy-Leblond, Jean-Marc},
  journal =       {Annales de l'I.H.P. Physique th{\'e}orique},
  number =        {1},
  pages =         {1-12},
  publisher =     {Gauthier-Villars},
  title =         {Une nouvelle limite non-relativiste du groupe de
                   {P}oincar{\'e}},
  volume =        {3},
  year =          {1965},
  language =      {fre},
  url =           {http://eudml.org/doc/75509},
}

@article{SenGupta1966OnAA,
  author =        {N. D. Sen Gupta},
  journal =       {Il Nuovo Cimento A (1965-1970)},
  pages =         {512-517},
  title =         {On an analogue of the {G}alilei group},
  volume =        {44},
  year =          {1966},
}

@article{Baiguera:2022lsw,
    author = "Baiguera, Stefano and Oling, Gerben and Sybesma, Watse and S\o{}gaard, Benjamin T.",
    title = "{Conformal Carroll scalars with boosts}",
    eprint = "2207.03468",
    archivePrefix = "arXiv",
    primaryClass = "hep-th",
    reportNumber = "NORDITA 2022-047",
    doi = "10.21468/SciPostPhys.14.4.086",
    journal = "SciPost Phys.",
    volume = "14",
    number = "4",
    pages = "086",
    year = "2023"
}

@article{Campoleoni:2022ebj,
    author = "Campoleoni, Andrea and Henneaux, Marc and Pekar, Simon and P\'erez, Alfredo and Salgado-Rebolledo, Patricio",
    title = "{Magnetic Carrollian gravity from the Carroll algebra}",
    eprint = "2207.14167",
    archivePrefix = "arXiv",
    primaryClass = "hep-th",
    doi = "10.1007/JHEP09(2022)127",
    journal = "JHEP",
    volume = "09",
    pages = "127",
    year = "2022"
}

@article{Bidussi:2021nmp,
    author = "Bidussi, Leo and Hartong, Jelle and Have, Emil and Musaeus, J\o{}rgen and Prohazka, Stefan",
    title = "{Fractons, dipole symmetries and curved spacetime}",
    eprint = "2111.03668",
    archivePrefix = "arXiv",
    primaryClass = "hep-th",
    doi = "10.21468/SciPostPhys.12.6.205",
    journal = "SciPost Phys.",
    volume = "12",
    number = "6",
    pages = "205",
    year = "2022"
}

@article{Hartong:2015xda,
    author = "Hartong, Jelle",
    title = "{Gauging the Carroll Algebra and Ultra-Relativistic Gravity}",
    eprint = "1505.05011",
    archivePrefix = "arXiv",
    primaryClass = "hep-th",
    doi = "10.1007/JHEP08(2015)069",
    journal = "JHEP",
    volume = "08",
    pages = "069",
    year = "2015"
}

@article{Bagchi:2020ats,
    author = "Bagchi, Arjun and Banerjee, Aritra and Chakrabortty, Shankhadeep",
    title = "{Rindler Physics on the String Worldsheet}",
    eprint = "2009.01408",
    archivePrefix = "arXiv",
    primaryClass = "hep-th",
    doi = "10.1103/PhysRevLett.126.031601",
    journal = "Phys. Rev. Lett.",
    volume = "126",
    number = "3",
    pages = "031601",
    year = "2021"
}

@article{Bagchi:2015nca,
    author = "Bagchi, Arjun and Chakrabortty, Shankhadeep and Parekh, Pulastya",
    title = "{Tensionless Strings from Worldsheet Symmetries}",
    eprint = "1507.04361",
    archivePrefix = "arXiv",
    primaryClass = "hep-th",
    reportNumber = "MIT-CTP-4690",
    doi = "10.1007/JHEP01(2016)158",
    journal = "JHEP",
    volume = "01",
    pages = "158",
    year = "2016"
}

@article{Grumiller:2020elf,
    author = "Grumiller, Daniel and Hartong, Jelle and Prohazka, Stefan and Salzer, Jakob",
    title = "{Limits of JT gravity}",
    eprint = "2011.13870",
    archivePrefix = "arXiv",
    primaryClass = "hep-th",
    reportNumber = "TUW--20--05",
    doi = "10.1007/JHEP02(2021)134",
    journal = "JHEP",
    volume = "02",
    pages = "134",
    year = "2021"
}

@article{deBoer:2021jej,
    author = "de Boer, Jan and Hartong, Jelle and Obers, Niels A. and Sybesma, Watse and Vandoren, Stefan",
    title = "{Carroll Symmetry, Dark Energy and Inflation}",
    eprint = "2110.02319",
    archivePrefix = "arXiv",
    primaryClass = "hep-th",
    reportNumber = "NORDITA 2021-086",
    doi = "10.3389/fphy.2022.810405",
    journal = "Front. in Phys.",
    volume = "10",
    pages = "810405",
    year = "2022"
}

@article{Perez:2021abf,
    author = "P\'erez, Alfredo",
    title = "{Asymptotic symmetries in Carrollian theories of gravity}",
    eprint = "2110.15834",
    archivePrefix = "arXiv",
    primaryClass = "hep-th",
    reportNumber = "CECS-PHY-21/03",
    doi = "10.1007/JHEP12(2021)173",
    journal = "JHEP",
    volume = "12",
    pages = "173",
    year = "2021"
}

@article{Donnay:2022aba,
    author = "Donnay, Laura and Fiorucci, Adrien and Herfray, Yannick and Ruzziconi, Romain",
    title = "{Carrollian Perspective on Celestial Holography}",
    eprint = "2202.04702",
    archivePrefix = "arXiv",
    primaryClass = "hep-th",
    doi = "10.1103/PhysRevLett.129.071602",
    journal = "Phys. Rev. Lett.",
    volume = "129",
    number = "7",
    pages = "071602",
    year = "2022"
}

@article{Ciambelli:2018wre,
    author = "Ciambelli, Luca and Marteau, Charles and Petkou, Anastasios C. and Petropoulos, P. Marios and Siampos, Konstantinos",
    title = "{Flat holography and Carrollian fluids}",
    eprint = "1802.06809",
    archivePrefix = "arXiv",
    primaryClass = "hep-th",
    reportNumber = "CPHT-RR049.082017, CERN-TH-2017-229",
    doi = "10.1007/JHEP07(2018)165",
    journal = "JHEP",
    volume = "07",
    pages = "165",
    year = "2018"
}

@article{Barnich:2006av,
      author         = "Barnich, Glenn and Comp{\`e}re, Geoffrey",
      title          = "{Classical central extension for asymptotic symmetries at null infinity in three spacetime dimensions}",
      journal        = "Class.Quant.Grav.",
      volume         = "24",
      pages          = "F15-F23",
      doi            = "10.1088/0264-9381/24/5/F01, 10.1088/0264-9381/24/11/C01",
      year           = "2007",
      eprint         = "gr-qc/0610130",
      archivePrefix  = "arXiv",
      primaryClass   = "gr-qc",
      reportNumber   = "ULB-TH-06-08",
      SLACcitation   = "%%CITATION = GR-QC/0610130;%%",
}

@Article{Regge:1974zd,
     author    = "Regge, Tullio and Teitelboim, Claudio",
     title     = "ROLE OF SURFACE INTEGRALS IN THE {H}AMILTONIAN FORMULATION OF
                  GENERAL RELATIVITY",
     journal   = "Ann. Phys.",
     volume    = "88",
     year      = "1974",
     pages     = "286",
     SLACcitation  = "%%CITATION = APNYA,88,286;%%"
}

@Book{waldgeneral,
  author = 	 {Robert M. Wald},
  title = 	 "{General Relativity}",
  publisher = 	 {The University of Chicago Press},
  year = 	 1984
}

\end{document}